\documentclass[11pt,a4paper]{article}

\usepackage{jheppub}
\usepackage{graphicx}
\usepackage{xcolor}
\usepackage{mathrsfs,mathtools}
\usepackage{physics,amssymb}
\usepackage{siunitx}
\usepackage{bm}
\usepackage{braket}
\usepackage{cases}
\usepackage{comment}
\usepackage{soul}
\usepackage{cancel}
\usepackage[utf8]{inputenc}
\usepackage{url}
\usepackage{xspace}
\usepackage{acronym}
\usepackage{tikz-feynhand}
\usepackage{subcaption}
\usepackage{aligned-overset}
\usepackage{microtype}

\hypersetup{colorlinks=true
,urlcolor=DARKBLUE
,anchorcolor=DARKBLUE
,citecolor=DARKBLUE
,filecolor=DARKBLUE
,linkcolor=DARKBLUE
,menucolor=DARKBLUE
,linktocpage=true
,pdfproducer=medialab
}

\newcommand{\ee}{\mathrm{e}}

\newcommand{\ns}{n_{\mathrm{s}}}

\newcommand{\umin}{\mathrm{min}}
\newcommand{\umax}{\mathrm{max}}
\newcommand{\enh}{\mathrm{enh}}
\newcommand{\SR}{\mathrm{SR}}
\newcommand{\fNL}{f_\mathrm{NL}}
\newcommand{\ph}{\mathrm{ph}}
\newcommand{\ren}{\mathrm{ren}}
\newcommand{\tad}{\mathrm{tad}}
\newcommand{\bw}{\mathrm{bw}}
\newcommand{\cl}{\mathrm{cl}}

\newcommand{\calA}{\mathcal{A}}
\newcommand{\calB}{\mathcal{B}}

\newcommand{\uc}{\mathrm{c}}

\newcommand{\ue}{\mathrm{e}}
\newcommand{\uf}{\mathrm{f}}

\newcommand{\uI}{\mathrm{I}}

\newcommand{\bmk}{\bm{k}}
\newcommand{\uL}{\mathrm{L}}

\newcommand{\calN}{\mathcal{N}}

\newcommand{\calO}{\mathcal{O}}
\newcommand{\scrO}{\mathscr{O}}
\newcommand{\calP}{\mathcal{P}}

\newcommand{\bmp}{\bm{p}}
\newcommand{\bmq}{\bm{q}}

\newcommand{\uS}{\mathrm{S}}
\newcommand{\us}{\mathrm{s}}
\newcommand{\bmx}{\bm{x}}

\newcommand{\beae}[1]{\begin{equation}\begin{aligned} #1 \end{aligned}\end{equation}}

\newcommand{\bae}[1]{\begin{align} #1 \end{align}}
\newcommand{\bce}[1]{\begin{cases} #1 \end{cases}}

\newcommand{\bme}[1]{\begin{multline} #1 \end{multline}}
\newcommand{\bmte}[1]{\begin{multlined}[t] #1 \end{multlined}}
\newcommand{\bmbe}[1]{\begin{multlined}[b] #1 \end{multlined}}

\definecolor{MONZA}{HTML}{CF000F}
\definecolor{DARKBLUE}{HTML}{00008b}
\definecolor{DARKMAGENTA}{HTML}{8b008b}
\definecolor{PURPLE}{rgb}{0.4 ,0, 0.85}

\acrodef{PBH}{primordial black hole}
\acrodef{CMB}{cosmic microwave background}
\acrodef{1PI}{one-particle irreducible}
\acrodef{NG}{Nambu--Goldstone}
\acrodef{UV}{ultraviolet}
\acrodef{IR}{infrared}
\acrodef{EFT}{effective field theory}
\acrodef{LHC}{Large Hadron Collider}
\acrodef{SR}{slow-roll}
\acrodef{USR}{ultra-slow-roll}

\title{Cancellation of quantum corrections on the soft curvature perturbations}
\date{\today}

\author[a, b]{Yuichiro Tada,}  
\author[c]{Takahiro Terada,}
\author[c]{and Junsei Tokuda}

\affiliation[a]{Institute for Advanced Research, Nagoya University,
Furo-cho Chikusa-ku, Nagoya 464-8601, Japan}
\affiliation[b]{Department of Physics, Nagoya University,
Furo-cho Chikusa-ku, Nagoya 464-8602, Japan}
\affiliation[c]{Particle Theory and Cosmology Group, Center for Theoretical Physics of the Universe, Institute for Basic Science (IBS),
Daejeon, 34126, Korea}

\emailAdd{tada.yuichiro.y8@f.mail.nagoya-u.ac.jp}
\emailAdd{takahiro.terada.hepc@gmail.com}
\emailAdd{jtokuda@ibs.re.kr}

\abstract{
We study the cancellation of quantum corrections on the superhorizon curvature perturbations from subhorizon physics beyond the single-clock inflation from the viewpoint of the cosmological soft theorem. As an example, we focus on the transient ultra-slow-roll inflation scenario and compute the one-loop quantum corrections to the power spectrum of curvature perturbations taking into account nontrivial surface terms in the action. We find that Maldacena's consistency relation is satisfied and guarantees the cancellation of contributions from the short-scale modes. As a corollary, primordial black hole production in single-field inflation scenarios is not excluded by perturbativity breakdown even for the sharp transition case in contrast to some recent claims in the literature. We also comment on the relation between the tadpole diagram in the in-in formalism and the shift of the elapsed time in the stochastic-$\delta N$ formalism. We find our argument is not directly generalisable to the tensor perturbations.  
}

\begin{document}
{\baselineskip0pt
\rightline{\baselineskip16pt\rm\vbox to-20pt{
           \hbox{CTPU-PTC-23-31}
\vss}}%
}

\maketitle
\flushbottom

\acresetall
\section{Introduction}

A hierarchy of scales allows one to integrate out \ac{UV} modes to obtain a low-energy \ac{EFT}.  For example, descriptions of the scattering of the Standard Model particles at the \acl{LHC} do not require treatment in quantum gravity. Similarly, chemical reactions can be understood without the knowledge of the Standard Model of particle physics. The essential underlying features are the scale separation and the decoupling of the \ac{UV} modes~\cite{Kazama:1981fx}.\footnote{
In quantum gravity and theories beyond local quantum field theories such as noncommutative field theory, the separation is not necessarily maintained, which is known as the \ac{UV}/\acs{IR} mixing~\cite{Minwalla:1999px}. 
We do not consider such effects as we focus on local quantum field theory without quantum gravity effects. 
} Applications of \acp{EFT} are quite successful in wide areas of physics.

Analogously, scale separation and decoupling of small-scale modes should be expected in quantum field theory in curved spacetime in the cosmological context. Indeed, one can formulate cosmological \acp{EFT} such as the \ac{EFT} of inflation~\cite{Cheung:2007st} and the \ac{EFT} of large-scale structure~\cite{Carrasco:2012cv} in the same manner as in the usual \ac{EFT} in flat spacetime. 
Somewhat relatedly, it is well known that the primordial curvature perturbations are conserved on superhorizon scales~\cite{Maldacena:2002vr, Lyth:2004gb, Langlois:2005ii, Langlois:2005qp} in adiabatic time evolution. They are not affected by subhorizon dynamics.  Moreover, it was shown in Refs.~\cite{Pimentel:2012tw, Senatore:2012ya, Assassi:2012et} that quantum corrections to the primordial curvature perturbations on superhorizon scales from much smaller scales are absent in single-clock inflation models.

Recently, these fundamental notions were challenged in Refs.~\cite{Cheng:2021lif, Kristiano:2022maq}, where large quantum corrections were found to the power spectrum of the primordial curvature perturbation on large scales from enhanced small-scale modes in \ac{SR} inflation with a transient period of \ac{USR} inflation~\cite{Inoue:2001zt, Tsamis:2003px, Kinney:2005vj}. During the \ac{USR} period, the inflaton field velocity in addition to its field value becomes relevant for its dynamics, so the single-clock nature of the single-field \ac{SR} inflation is lost thereby evading the theorem~\cite{Pimentel:2012tw,Senatore:2012ya, Assassi:2012et} that forbids quantum corrections. Since the enhancement of small-scale curvature perturbations is typically employed for \ac{PBH} formation~\cite{Hawking:1971ei, Carr:1974nx, Carr:1975qj} (see also recent review articles~\cite{Carr:2020gox,Escriva:2022duf,Yoo:2022mzl,Carr:2023tpt}), it was further claimed in Ref.~\cite{Kristiano:2022maq} that \ac{PBH} formation in single-field inflation is ``ruled out'' because of violation of perturbativity.  Subsequently, a series of papers discussed this issue~\cite{Riotto:2023hoz, Choudhury:2023vuj, Choudhury:2023jlt, Kristiano:2023scm, Riotto:2023gpm, Choudhury:2023rks, Firouzjahi:2023aum, Motohashi:2023syh, Choudhury:2023hvf, Firouzjahi:2023ahg, Franciolini:2023lgy, Tasinato:2023ukp, Cheng:2023ikq, Fumagalli:2023hpa, Maity:2023qzw}.  

Our focus in this paper is the quantum corrections to the large-scale cosmological perturbations themselves rather than \acp{PBH}. 
While it is important to answer the question of whether \ac{PBH} formation in single-field inflation is possible without large backreactions to the large scale where the \ac{CMB} fluctuations are measured, discussions on this question so far largely depend on the results that the one-loop corrections to the \ac{CMB} scale can be sizable~\cite{Cheng:2021lif, Kristiano:2022maq, Riotto:2023hoz, Choudhury:2023vuj, Choudhury:2023jlt, Kristiano:2023scm, Choudhury:2023rks, Firouzjahi:2023aum, Franciolini:2023lgy, Cheng:2023ikq, Maity:2023qzw}.  However, the full one-loop calculation including all the relevant vertices has not yet been performed.  If the large contributions cancel so that there is no total large quantum correction from small-scale loops (even in the case of sharp \ac{SR}/\ac{USR}/\ac{SR} transitions) as suggested in Refs.~\cite{Tasinato:2023ukp, Fumagalli:2023hpa}, the threat to \ac{PBH} formation in single-field inflation disappears completely.  In particular, Ref.~\cite{Fumagalli:2023hpa} pointed out the importance of a total derivative term to see such a cancellation. We augment his calculation with another relevant surface term and emphasise the role of Maldacena's consistency relation~\cite{Maldacena:2002vr} along the line of Ref.~\cite{Pimentel:2012tw} to understand the fact that the quantum corrections vanish as a result of a cosmological soft theorem on the one-loop diagrams. 

In addition to the issue of the quantum corrections to the soft (large-scale) curvature perturbations from the small-scale loop effects, quantum corrections to the primordial power spectrum of gravitational waves on large-scales from small-scale loop effects were found in the same \ac{SR}/\ac{USR}/\ac{SR} scenario~\cite{Firouzjahi:2023btw} and in a phenomenological setup where a mode function of a spectator scalar field is enhanced on small scales during \ac{SR} inflation~\cite{Ota:2022hvh, Ota:2022xni}. Although these quantum corrections would not violate the perturbativity on \ac{CMB} scales (at least when we restrict the mode function of the spectator field to be of the Bogoliubov transformation form~\cite{Ota:2022xni}), it is in serious tension with the notion of decoupling. We will briefly discuss this issue as well.  

Another purpose of this paper is to point out the relation between the in-in (Schwinger--Keldysh) formalism~\cite{Schwinger:1960qe, Keldysh:1964ud, Feynman:1963fq} (for review, see Refs.~\cite{Kamenev:2009jj, Chen:2017ryl} and references therein) and the stochastic-$\delta N$ formalism (see, e.g., Refs.~\cite{Fujita:2013cna,Fujita:2014tja,Vennin:2015hra,Ando:2020fjm,Tada:2021zzj}) for the calculation of cosmological correlation functions. The stochastic formalism (see Refs.~\cite{Starobinsky:1982ee,Starobinsky:1986fx,Nambu:1987ef,Nambu:1988je,Kandrup:1988sc,Nakao:1988yi,Nambu:1989uf,Mollerach:1990zf,Linde:1993xx,Starobinsky:1994bd} for the first papers on the subject) is the \ac{EFT} for matter fields (such as the inflaton) coarse-grained on a superhorizon scale, called the \ac{IR} mode. There, the \ac{IR} mode is interpreted as a (non-quantum) stochastic variable (or a Brownian motion), governed by the effective action improved by the \ac{UV} loops.
Combining it with the $\delta N$ formalism~\cite{Salopek:1990jq,Sasaki:1995aw,Sasaki:1998ug,Lyth:2004gb}, one can calculate the statistical properties of the curvature perturbations in the stochastic formalism. Similarly to the ordinary \ac{EFT} approach, the stochastic computation should be consistent with the direct calculation in the in-in approach. We point out that the shift of the average elapsed time of inflation can be a correction on large-scale perturbations from small-scale ones in the stochastic approach and it is understood as a tadpole contribution in the in-in formalism.

The structure of the paper is as follows.  In Sec.~\ref{sec: soft}, we review the consistency relation and discuss why the cancellation of one-loop corrections is generically expected.  As a concrete and important example, we consider the transient \ac{USR} inflation in Sec.~\ref{sec: example}. We confirm the consistency relation in this setup and use the results in the calculation of the one-loop corrections to the power spectrum of the large-scale curvature perturbations. We find the cancellation of dominant contributions, so there is no significant one-loop correction to the power spectrum of the curvature perturbations. In Sec.~\ref{sec: tadpole}, we discuss the tadpole diagram and its relation to the counterpart in the stochastic-$\delta N$ formalism.  We argue that the tadpole diagram, which can be absorbed by the redefinition of the scale factor, is the only possible nonzero contribution to the quantum corrections of the power spectrum of the curvature perturbations in the soft limit. We briefly discuss the tensor power spectrum in Sec.~\ref{sec: tensor}. Sec.~\ref{sec: conclusion} is devoted to our conclusions. Throughout the paper, we take the Planck unit $c = \hbar = 8\pi G = 1$.

\section{Soft theorem on the cosmological loop corrections \label{sec: soft}}

In this section, we schematically discuss the possible form of the self-energy of the soft curvature perturbation from the viewpoint of the cosmological soft theorem. We particularly focus on the corrections from \ac{UV} fluctuations whose wave numbers are much larger than that of the external lines.

Symmetries of the theory restrict the allowed form of Green's functions, \ac{1PI} vertices, or their generating functional, via the Ward--Takahashi identity~\cite{Ward:1950xp, Takahashi:1957xn}. 
In cosmology, the so-called Maldacena's consistency relation~\cite{Maldacena:2002vr} (see also Refs.~\cite{Chen:2006dfn,Li:2008gg,Leblond:2010yq,Seery:2008ax} and Refs.~\cite{Bravo:2017wyw, Finelli:2017fml} for its generalisation to $N$-point functions and to the \ac{USR} models, respectively) is understood as an example of the Ward--Takahashi identity on correlation functions of cosmological perturbations including at least one soft external/internal momentum~\cite{Creminelli:2004yq,Cheung:2007sv,Creminelli:2011rh}. 
The curvature perturbation\footnote{Throughout this work, we consider the genuine curvature perturbation $\zeta$ rather than quantities like $\zeta_n$ related to the original $\zeta$ by field redefinitions~\cite{Maldacena:2002vr}.} $\zeta$ is regarded as the \ac{NG} boson associated with the dilatation~\cite{Creminelli:2012ed,Hinterbichler:2012nm,Assassi:2012zq, Hinterbichler:2013dpa,Goldberger:2013rsa,Kundu:2014gxa,Kundu:2015xta}. 
The squeezed limit of the bispectra in the adiabatic universe for example is then given by the (Fourier-space) dilatation of the power spectrum of the other operator $\hat{\calO}$ as~\cite{Maldacena:2002vr,Creminelli:2011rh,Pajer:2013ana}
\bae{\label{eq: Maldacena's CR}
	B_{\zeta\calO\calO}(k_1,k_2,k_3)\underset{k_1\ll k_2\sim k_3}{\to} - P_\zeta(k_1)P_\calO(k_\uS)\dv{\ln\calP_\calO(k_\uS)}{\ln k_\uS}+\scrO\pqty{\pqty{\frac{k_1}{k_\uS}}^2},
}
as the \ac{NG} soft theorem, where $k_\uS\coloneqq\abs{(\bmk_2-\bmk_3)/2}$ and the power spectrum and bispectrum are defined by
\beae{
	&\braket{\hat{\calO}_1(\bmk_1)\hat{\calO}_2(\bmk_2)}=(2\pi)^3\delta^{(3)}(\bmk_1+\bmk_2)P_{\calO_1\calO_2}(k_1), \\
	&\braket{\hat{\calO}_1(\bmk_1)\hat{\calO}_2(\bmk_2)\hat{\calO}_3(\bmk_3)}=(2\pi)^3\delta^{(3)}(\bmk_1+\bmk_2+\bmk_3)B_{\calO_1\calO_2\calO_3}(k_1,k_2,k_3), \\
	&\calP_{\calO_1\calO_2}(k)=\frac{k^3}{2\pi^2}P_{\calO_1\calO_2}(k) \qc
	P_\calO(k)=P_{\calO\calO}(k) \qc \calP_\calO(k)=\calP_{\calO\calO}(k).
}
It means that the short-long correlation is dominated by the ``artefact" due to the modulation of the local spatial metric, which can be renormalised into the redefinition of the local metric~\cite{Urakawa:2009my,Urakawa:2009gb,Tanaka:2011aj,Pajer:2013ana,Bartolo:2015qva,Dai:2015jaa,dePutter:2015vga,Tada:2016pmk,Suyama:2020akr,Suyama:2021adn}.
Therefore, the satisfaction of Maldacena's consistency relation implies no ``physical" correlation between the short- and long-wavelength modes.
We numerically confirm the consistency relation beyond the \ac{SR} inflation in Sec.~\ref{sec: example}. 
Does this soft theorem have an implication for self-energy?

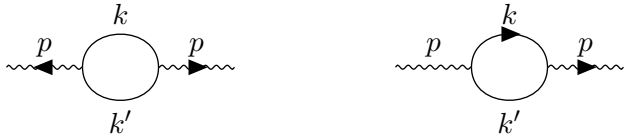
\begin{figure}
	\centering
	\begin{tabular}{c}
		\begin{minipage}{0.33\hsize}
			\centering
			\begin{tikzpicture}
            	\begin{feynhand}
					\vertex (a) at (-1.5,0);
					\vertex (b) at (-0.5,0);
					\vertex (c) at (0.5,0);
					\vertex (d) at (1.5,0);
					\propag[chabos] (b) to [edge label'=$p$] (a);
					\propag[plain] (b) to [out=90, in=90, looseness=1.5, edge label=$k$] (c);
					\propag[plain] (b) to [out=-90, in=-90, looseness=1.5, edge label'=$k'$] (c);
					\propag[chabos] (c) to [edge label=$p$] (d);
				\end{feynhand}
			\end{tikzpicture}
		\end{minipage}
		\begin{minipage}{0.33\hsize}
			\centering
			\begin{tikzpicture}
            	\begin{feynhand}
					\vertex (a) at (-1.5,0);
					\vertex (b) at (-0.5,0);
					\vertex (c) at (0.5,0);
					\vertex (d) at (1.5,0);
					\propag[bos] (b) to [edge label'=$p$] (a);
					\propag[fermion] (b) to [out=90, in=90, looseness=1.5, edge label=$k$] (c);
					\propag[plain] (b) to [out=-90, in=-90, looseness=1.5, edge label'=$k'$] (c);
					\propag[chabos] (c) to [edge label=$p$] (d);
				\end{feynhand}
			\end{tikzpicture}
		\end{minipage}
	\end{tabular}
    \caption{Two representative diagrams of the one-loop corrections on the propagator of cosmological perturbations through the cubic interactions. The wave lines represent the external curvature perturbation, while the plain lines are internal particles also for which we assume the curvature perturbation in this paper. Lines without the arrow are the statistical propagators and ones with the arrow indicate the retarded propagators (see Sec.~\ref{sec: example} and also Ref.~\cite{Motohashi:2023syh}). 
    \emph{Left}: the so-called cut-in-the-middle diagram representing the \emph{induction} of long-wavelength modes by short-wavelength modes through the second-order effect such as the induced gravitational waves~\cite{1967PThPh..37..831T, Matarrese:1993zf, Matarrese:1997ay, Ananda:2006af, Baumann:2007zm}. \emph{Right}: the cut-in-the-side diagram which causes the effective mass correction. It has the potential ability to alter even the soft propagator.}
    \label{fig: self energy}
\end{figure}

In the Schwinger--Keldysh formalism, the (possible) dominant contributions for the self-energy via the cubic interaction\footnote{In this work, we mainly focus on the cubic interaction to see the relation between the self-energy and the squeezed bispectrum. Though a similar discussion is expected for the quartic interaction, we leave it for future works. See Refs.~\cite{Firouzjahi:2023aum, Maity:2023qzw} for some discussions on quartic interactions.  See also Sec.~\ref{sec: tadpole} for the tadpole contribution.} are schematically classified into the two Feynman diagrams shown in Fig.~\ref{fig: self energy} (see Sec.~\ref{sec: example} or Ref.~\cite{Motohashi:2023syh} for the details of the Feynman rule).
The left one called the cut-in-the-middle diagram~\cite{Pimentel:2012tw} represents the induction of long-wavelength modes by short-wavelength modes through the second-order effect such as the induced gravitational waves~\cite{1967PThPh..37..831T, Matarrese:1993zf, Matarrese:1997ay, Ananda:2006af, Baumann:2007zm} (see Ref.~\cite{Domenech:2021ztg} for a review). 
This diagram is independent of $p$ because the retarded propagator (the line with the arrow) is independent of $p$ at the leading order.\footnote{Explicit expressions of the retarded propagator on the superhorizon scales are provided in Appendix~\ref{sec:calculations}.}
As the corresponding dimensionless power spectrum includes the factor of $p^3$ via the Fourier-space volume factor, this diagram is suppressed in the soft limit $p\to0$, whose feature has been accepted in the literature.

On the other hand, the right cut-in-the-side diagram~\cite{Pimentel:2012tw} is understood as the effective mass correction, which is proportional to the statistical propagator (the line without the arrow) of $p$. It is proportional to $p^{-3}$ and hence this diagram may cause a scale-invariant correction on the soft propagator. In terms of the power spectrum, the diagram formally has a structure of $C(k)P_{\zeta}(p)P_{\calO}(k)$ with a coefficient function $C(k)$ of $k$. If the short-wavelength perturbations $P_{\calO}(k)$ are enhanced for \ac{PBH} formation for example, it could potentially cause a significant correction even on the \acs{CMB}-scale perturbation, which is the reason why the loop correction has recently attracted much attention.

\begin{figure}
    \centering
    \begin{tikzpicture}
        \begin{feynhand}
            \vertex (a) at (-1.5,1);
            \vertex (b) at (0.5,1);
            \vertex (c) at (1.5,1);
            \vertex (d) at (0,-1);
            \propag[bos] (d) to [out=135, in=-90, edge label=$p$] (a);
            \propag[fermion] (d) to [edge label=$k$] (b);
            \propag[plain] (d) to [out=45, in=-90, edge label'=$k'$] (c);
        \end{feynhand}
    \end{tikzpicture}
    \caption{Schematic representation of the Feynman diagram for the squeezed bispectrum $B_{\zeta\calO\calO}(p,k,k')$.}
    \label{fig: bispectrum}
\end{figure}
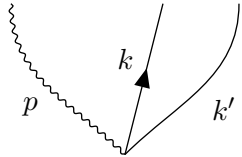

However, one should note that the coefficient $C(k)$ is not necessarily positive- or negative-definite and there can be cancellation in the integration over the loop momentum $k$.
In fact, the cut-in-the-side diagram includes the bispectrum one shown in Fig.~\ref{fig: bispectrum} as a subdiagram.
Therefore, if the squeezed bispectrum satisfies Maldacena's consistency relation~\eqref{eq: Maldacena's CR}, the diagram contribution should be summarised as $cP_\zeta(p)P_\calO(k)\dv{\ln\calP_\calO(k)}{\ln k}$ with a constant $c$.
The loop contribution is hence proportional to
\bae{
    P_{\zeta}(p)\int_{\ln k_\umin}^{\ln k_\umax}\frac{k^3}{2\pi^2}P_\calO(k)\dv{\ln\calP_\calO(k)}{\ln k} \dd{\ln k} = P_{\zeta}(p)\pqty{\calP_\calO(k_\umax)-\calP_\calO(k_\umin)}.
}
In principle, the integration should be taken over the whole momentum as $k_\umin\to0$ and $k_\umax\to\infty$.
In order to utilise the squeezed bispectrum, we however set $k_\umin\gg p$ and neglect the contribution of $k<k_\umin$, which is enough for our purpose. Nevertheless, $k_\umin$ should be sufficiently smaller than the enhancement scale $k_\enh$ so that the corresponding power spectrum is much smaller than the enhanced one: $\calP_\calO(k_\umin)\ll\calP_\calO(k_\enh)$.
The \ac{UV} contribution $\calP_\calO(k_\umax\to\infty)$ can be dropped by the $i\varepsilon$ prescription, i.e., by rotating the time axis as $\tau\to(1+i\varepsilon)\tau$ with a small positive parameter $\varepsilon$ and replacing the \ac{UV} mode function as $\ee^{-ik\tau}\to\ee^{-ik\tau+\varepsilon k\tau}$. Indeed, this procedure is necessary to correctly evaluate the correlation functions in the vacuum state of the interacting theory in the infinite past. 

As a result, if Maldacena's consistency relation holds for the squeezed bispectrum, the one-loop correction on the soft propagator from the cut-in-the-side diagram is summarised as $\sim P_{\zeta}(p)\calP_\calO(k_\umin)$, which is small enough compared with the tree-level result $P_{\zeta}(p)$.
It should be emphasised that the correction is independent of the enhanced power $\calP_\calO(k_\enh)$ and the enhancement of the small-scale power spectrum does not spoil the perturbativity of the large-scale perturbation as long as Maldacena's consistency relation holds true.
In the next section, we confirm this speculation in a specific inflation model: transient \ac{USR} inflation.

\section{Example: transient \acf{USR} inflation}\label{sec: example}

We show the concrete one-loop calculation in a specific inflation model: the transient \ac{USR} inflation considered in Refs.~\cite{Kristiano:2022maq,Kristiano:2023scm}.
There, the second \ac{SR} parameter $\eta=\dot{\epsilon}/(\epsilon H)$, where $\epsilon=-\dot{H}/H^2$ is the first \ac{SR} parameter, shows sharp transitions at specific conformal time $\tau_\us$ and $\tau_\ue$ as
\bae{
	\eta=-6\Theta(\tau-\tau_\us)\Theta(\tau_\ue-\tau),
}
so that the inflation dynamics is given by the \ac{SR} one with $\eta=0$ for $\tau<\tau_\us$ or $\tau>\tau_\ue$ while it is in the \ac{USR} phase with $\eta=-6$ during $\tau_\us\leq\tau\leq\tau_\ue$.
The first \ac{SR} parameter is solved at the leading-order \ac{SR} approximation as
\bae{
    \epsilon(\tau)=\bce{
    \epsilon_\SR, & \tau<\tau_\us, \\
    \epsilon_\SR\pqty{\frac{\tau}{\tau_\us}}^6, & \tau_\us\leq\tau\leq\tau_\ue, \\
    \epsilon_\SR\pqty{\frac{\tau_\ue}{\tau_\us}}^6, & \tau_\ue<\tau,
    }
}
with a certain initial value $\epsilon_\SR$.

\begin{figure}
    \centering
    \begin{tabular}{c}
        \begin{minipage}{0.5\hsize}
            \centering
            \includegraphics[width=0.95\hsize]{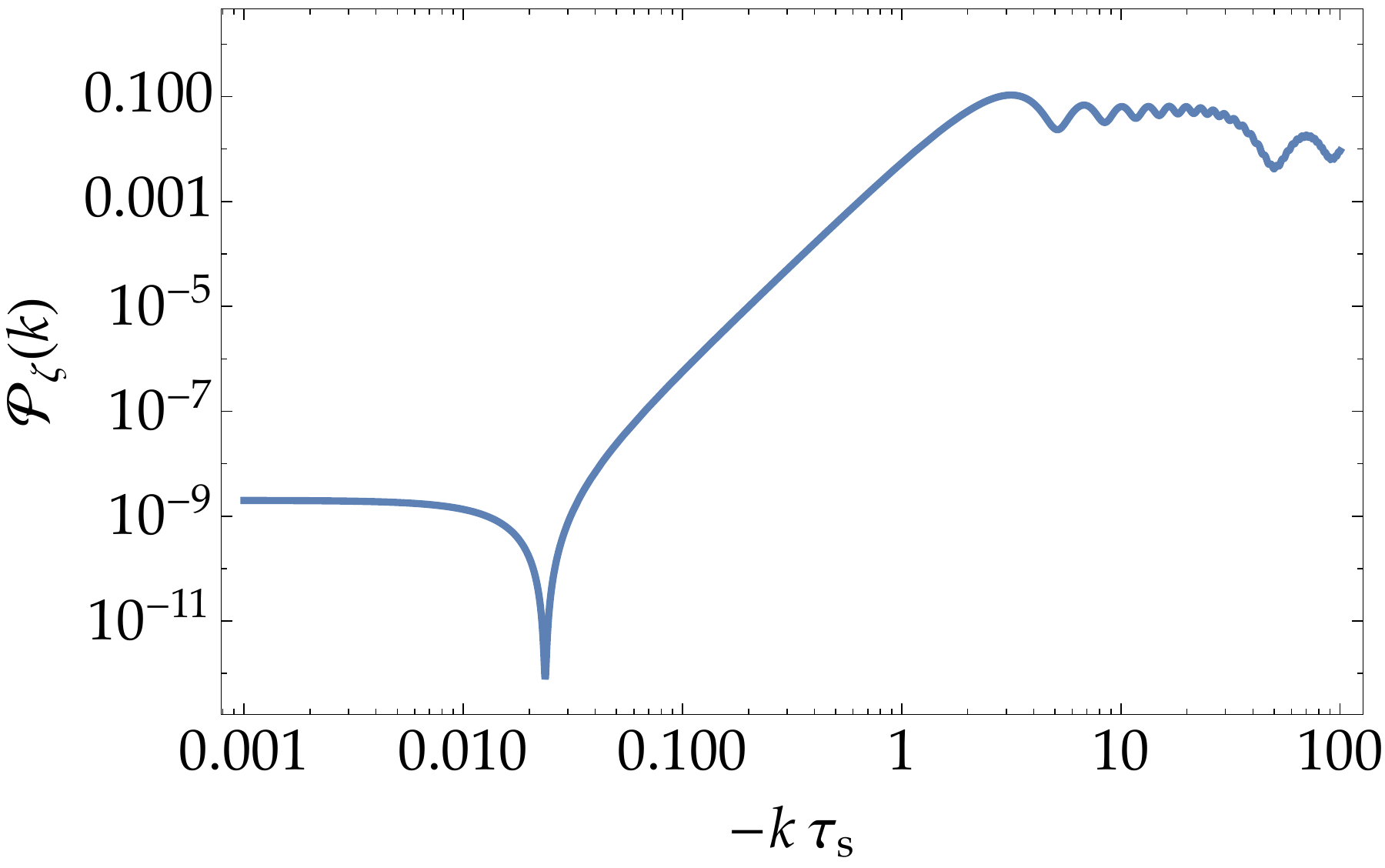}
        \end{minipage}
        \begin{minipage}{0.5\hsize}
            \centering
            \includegraphics[width=0.95\hsize]{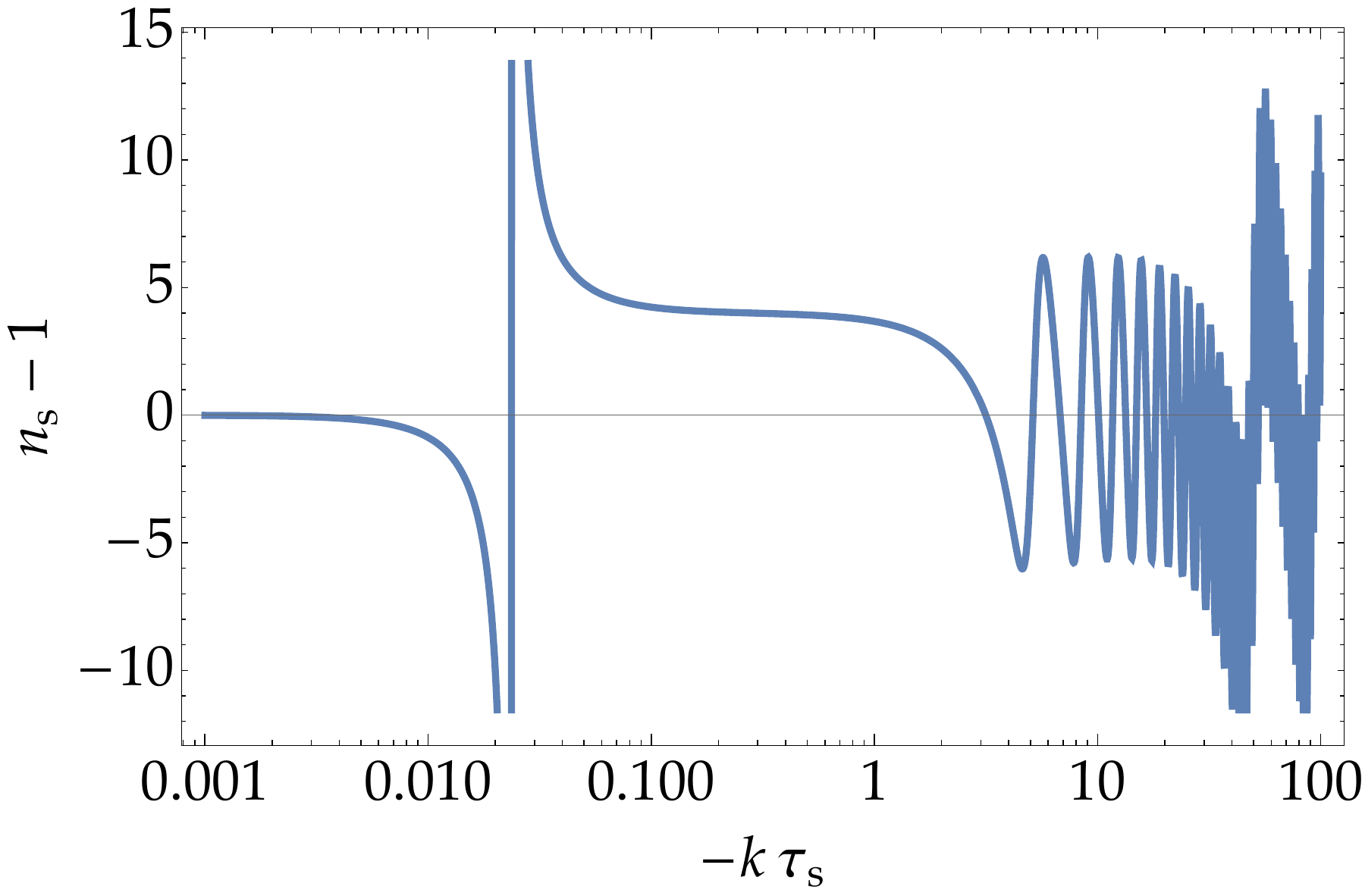}
        \end{minipage}
    \end{tabular}
    \caption{Example power spectrum $\calP_\zeta$ (left) and spectral index $\ns$ (right) for $H=\frac{\pi}{25000\sqrt{10}}$, $\epsilon_\SR=10^{-2}$, and $\frac{\tau_\us}{\tau_\ue}=5^{1/6}\times10$.}
    \label{fig: calP and ns SUS}
\end{figure}

In each phase, the formal solution of the Mukhanov--Sasaki equation for the mode function of the curvature perturbation is given by 
\bae{\label{eq: formal mode function}
    \zeta_k(\tau)=\frac{iH}{2\sqrt{\epsilon(\tau)}}\frac{1}{k^{3/2}}\bqty{\calA_k\ee^{-ik\tau}(1+ik\tau)-\calB_k\ee^{ik\tau}(1-ik\tau)}.
}
The initial condition is given by the Bunch--Davies vacuum $\calA_k=1$ and $\calB_k=0$ in the first \ac{SR} phase:
\bae{
    \zeta_k(\tau)=\zeta_k^{\SR1}(\tau)\coloneqq\frac{iH}{2\sqrt{\epsilon_\SR}k^{3/2}}\ee^{-ik\tau}(1+ik\tau) \qc
    \tau\leq\tau_\us.
}
Later, the solution is determined by the junction conditions $\zeta(\tau_{\us/\ue}-0)=\zeta(\tau_{\us/\ue}+0)$ and $\zeta'(\tau_{\us/\ue}-0)=\zeta'(\tau_{\us/\ue}+0)$ at the transitions $\tau_\us$ and $\tau_\ue$. The solution is given by~\cite{Kristiano:2023scm} 
\begin{align}
    \mathcal{A}_k = 1 + \frac{3 i (1 + (k \tau_\us)^2)}{2 (k \tau_\us)^3} \qc
    \mathcal{B}_k = - \frac{3 i (- i + k \tau_\us)^2 \ee^{- 2i k \tau_\us}}{2 (k \tau_\us)^3},
\end{align}
during the transient USR period ($\tau_\text{s} < \tau \leq \tau_\text{e}$) and 
\beae{
    &\mathcal{A}_k = \bmte{\frac{1}{{4 k^6 (\tau_\us \tau_\ue)^3}}\biggl[-9 (i + k \tau_\ue)^2(-i + k \tau_\us)^2 \ee^{2 i k (\tau_\ue-\tau_\us)} \\
    + \left(-3i + (-3i + 2k \tau_\ue)(k\tau_\ue)^2\right) \left( 3 i + (3i + 2 k \tau_\us)(k \tau_\us)^2 \right)\biggr],} \\
    &\mathcal{B}_k = \bmte{\frac{3i}{{4 k^6 (\tau_\us \tau_\ue)^3}}\biggl[ - (-i+k\tau_\us)^2 \left( 3 i + (3i + 2k \tau_\ue)(k \tau_\ue)^2\right)\ee^{-2i k \tau_\us} \\
    + (-i + k \tau_\ue)^2 (3i + (3i + 2k \tau_\us)(k\tau_\us)^2) \ee^{-2i k \tau_\ue}\biggr],}
}
in the second \ac{SR} period ($\tau > \tau_\ue$). 
With these expressions, we see explicitly that $|\zeta_k(\tau)|^2$ decreases exponentially as a function of $k$ upon the contour deformation $\tau \to (1 + i \varepsilon) \tau$. This fact is important when we discuss the \ac{UV} behaviour of the power spectrum $\mathcal{P}_\zeta (k)$. 

An example of the power spectrum $\calP_\zeta=\frac{k^3}{2\pi^2}\abs{\zeta_k}^2$ and its spectral index $\ns-1=\dv*{\ln\calP_\zeta}{\ln k}$ at late time $\tau\to0$ is shown in Fig.~\ref{fig: calP and ns SUS}. One sees the amplification of the curvature perturbation due to the transient \ac{USR} phase.

\subsection{Consistency relation on the squeezed bispectrum}

\begin{figure}
	\centering
	\begin{tabular}{c}
		\begin{minipage}{0.33\hsize}
			\centering
            \begin{tikzpicture}
                \begin{feynhand}
                    \vertex[particle] (a) at (-1.5,0) {$x$};
                    \vertex[particle] (b) at (1.5,0) {$x^\prime$};
                    \propag[plain] (a) to (b);
                \end{feynhand}
            \end{tikzpicture}
            \subcaption{$G_{\uc\uc}(x,x^\prime)=G_{\uc\uc}(x',x)$}
		\end{minipage}
		\begin{minipage}{0.33\hsize}
            \centering
            \begin{tikzpicture}
                \begin{feynhand}
                    \vertex[particle] (a) at (-1.5,0) {$x$};
                    \vertex[particle] (b) at (1.5,0) {$x^\prime$};
                    \propag[fermion] (b) to (a);
                \end{feynhand}
            \end{tikzpicture}
            \subcaption{$G_{\uc\Delta}(x,x^\prime)=G_{\Delta\uc}(x',x)$}
        \end{minipage} \\ \\
        \begin{minipage}{0.33\hsize}
            \centering
            \begin{tikzpicture}
                \begin{feynhand}
                    \vertex[particle] (a) at (-1.5,0) {$x$};
                    \vertex (b) at (0,0);
                    \vertex[particle] (c) at (1.5,0) {$x^\prime$};
                    \propag[plain] (a) to (b);
                    \propag[sca] (b) to (c);
                \end{feynhand}
            \end{tikzpicture}
            \subcaption{$G_{\uc\bar{\uc}}(x,x^\prime)=G_{\bar{\uc}\uc}(x',x)$}
        \end{minipage}
        \begin{minipage}{0.33\hsize}
            \centering
            \begin{tikzpicture}
                \begin{feynhand}
                    \vertex[particle] (a) at (-1.5,0) {$x$};
                    \vertex (b) at (0.5,0);
                    \vertex[particle] (c) at (1.5,0) {$x^\prime$};
                    \propag[fermion] (b) to (a);
                    \propag[sca] (b) to (c);
                \end{feynhand}
            \end{tikzpicture}
            \subcaption{$G_{\uc\bar{\Delta}}(x,x^\prime)=G_{\bar{\Delta}\uc}(x',x)$}
        \end{minipage}
        \begin{minipage}{0.33\hsize}
            \centering
            \begin{tikzpicture}
                \begin{feynhand}
                    \vertex[particle] (a) at (-1.5,0) {$x$};
                    \vertex (b) at (-0.5,0);
                    \vertex[particle] (c) at (1.5,0) {$x^\prime$};
                    \propag[sca] (a) to (b);
                    \propag[fermion] (c) to (b);
                \end{feynhand}
            \end{tikzpicture}
            \subcaption{$G_{\bar{\uc}\Delta}(x,x^\prime)=G_{\Delta\bar{\uc}}(x',x)$}
        \end{minipage} \\ \\
        \begin{minipage}{0.33\hsize}
            \centering
            \begin{tikzpicture}
                \begin{feynhand}
                    \vertex[particle] (a) at (-1.5,0) {$x$};
                    \vertex[particle] (b) at (1.5,0) {$x^\prime$};
                    \propag[sca] (a) to (b);
                \end{feynhand}
            \end{tikzpicture}
            \subcaption{$G_{\bar{\uc}\bar{\uc}}(x,x^\prime)=G_{\bar{\uc}\bar{\uc}}(x',x)$}
        \end{minipage}
        \begin{minipage}{0.33\hsize}
            \centering
            \begin{tikzpicture}
                \begin{feynhand}
                    \vertex[particle] (a) at (-1.5,0) {$x$};
                    \vertex[particle] (b) at (1.5,0) {$x^\prime$};
                    \propag[chasca] (b) to (a);
                \end{feynhand}
            \end{tikzpicture}
            \subcaption{$G_{\bar{\uc}\bar{\Delta}}(x,x^\prime)=G_{\bar{\Delta}\bar{\uc}}(x',x)$}
        \end{minipage}
	\end{tabular}
	\caption{Diagrams for propagators.}
	\label{fig: propagator}
\end{figure}
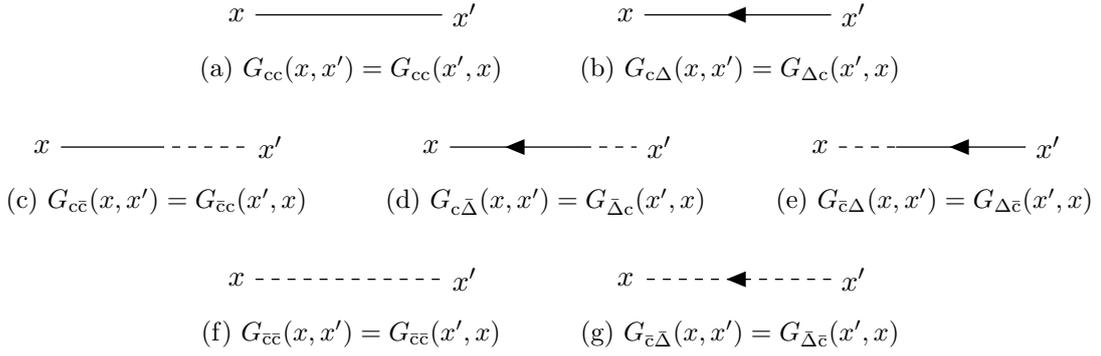

Let us first check if Maldacena's consistency relation is satisfied between the \ac{CMB}- and \ac{PBH}-scale perturbations, reviewing the Feynman diagrammatic approach in the Schwinger--Keldysh formalism. See Ref.~\cite{Motohashi:2023syh} for the detailed derivation.
The propagator is defined by the two-point function time-ordered along the closed-time path (denoted by $T_C$) as
\beae{
	&G_{\alpha\beta}(x,x')=\braket{T_C\hat{\zeta}_{\alpha\uI}(x)\hat{\zeta}_{\beta\uI}(x')} \qc 
	G_{\bar{\alpha}\bar{\beta}}(x,x')=\braket{T_C\hat{\zeta}'_{\alpha\uI}(x)\hat{\zeta}'_{\beta\uI}(x')}, \\
	&G_{\bar{\alpha}\beta}(x,x')=\braket{T_C\hat{\zeta}'_{\alpha\uI}(x)\hat{\zeta}_{\beta\uI}(x')} \qc
	G_{\alpha\bar{\beta}}(x,x')=\braket{T_C\hat{\zeta}_{\alpha\uI}(x)\hat{\zeta}'_{\beta\uI}(x')}.
}
Here, the indices $\alpha$ and $\beta$ label the Keldysh basis operators $\hat{\zeta}_\uc$ or $\hat{\zeta}_\Delta$, the subscript $\uI$ indicates that the operator is in the interaction picture, and $\zeta'$ denotes the generalised momentum $\partial_\tau\zeta$.
We diagrammatically illustrate them by plane or dashed lines with or without arrows as summarised in Fig.~\ref{fig: propagator}.
Lines without arrows are called \emph{statistical} propagators and ones with arrows are \emph{retarded} (or \emph{advanced}) propagators.
In Fourier space, they are given by 
\bae{
	G_{\alpha\beta}(\tau,\tau^\prime;k)=\bce{
		\Re \left( \zeta_k(\tau)\zeta_k^*(\tau^\prime) \right), & (\alpha,\beta)=(\uc,\uc), \\
		2i\Theta(\tau-\tau^\prime)\Im \left( \zeta_k(\tau)\zeta_k^*(\tau^\prime) \right), & (\alpha,\beta)=(\uc,\Delta), \\
		-2i\Theta(\tau^\prime-\tau)\Im \left( \zeta_k(\tau)\zeta_k^*(\tau^\prime) \right), & (\alpha,\beta)=(\Delta,\uc), \\
		0, & (\alpha,\beta)=(\Delta,\Delta),
	}
}
and similar ones including generalised momenta. Here, the function $\zeta_k(\tau)$ should be understood as the mode function~\eqref{eq: formal mode function} given above. The step function $\Theta(z)$ is defined by\footnote{Note that we changed the definition of the step function from Ref.~\cite{Motohashi:2023syh} for simplicity. 
Accordingly, the apparent vertex values of the boundary terms are changed.
This is a matter of convention, and it does not affect physical results.}
\bae{
	\Theta(z)=\bce{
		1, & z\geq0, \\
		0, & z<0.
	}
}
One should note that the equal-time statistical propagator is equivalent to the (symmetric) power spectrum, $G_{\uc\uc}(\tau,\tau;k)=P_{\zeta}(\tau,k)$, $G_{\uc\bar{\uc}}(\tau,\tau;k)=(P_{\zeta\zeta'}(\tau,k)+P_{\zeta'\zeta}(\tau,k))/2$, and $G_{\bar{\uc}\bar{\uc}}(\tau,\tau;k)=P_{\zeta'}(\tau,k)$.
The equal-time retarded propagator for $\zeta$ and $\zeta'$ is determined by the Wronskian condition as $G_{\uc\bar{\Delta}}(\tau,\tau;k)=i/(2a^2\epsilon)$ independently of $k$. The equal-time retarded one for the same operator, $G_{\uc\Delta}(\tau,\tau;k)$ or $G_{\bar{\uc}\bar{\Delta}}(\tau,\tau;k)$, vanishes by definition.

\begin{figure}
	\centering
	\begin{tabular}{c}
		\begin{minipage}{0.33\hsize}
			\centering
			\begin{tikzpicture}
				\begin{feynhand}
					\vertex[particle] (a) at (0,1.5) {$\Delta$};
					\vertex[particle] (b) at (-1.3,-0.75) {$\uc$};
					\vertex[particle] (c) at (1.3,-0.75) {$\uc$};
					\vertex (d) at (0,0);
					\propag[plain] (a) to (d);
					\propag[plain] (b) to (d);
					\propag[sca] (c) to (d);
				\end{feynhand}
			\end{tikzpicture}
		\end{minipage}
		\begin{minipage}{0.33\hsize}
			\centering
			\begin{tikzpicture}
				\begin{feynhand}
					\vertex[particle] (a) at (0,1.5) {$\Delta$};
					\vertex[particle] (b) at (-1.3,-0.75) {$\uc$};
					\vertex[particle] (c) at (1.3,-0.75) {$\uc$};
					\vertex (d) at (0,0);
					\propag[sca] (a) to (d);
					\propag[plain] (b) to (d);
					\propag[plain] (c) to (d);
				\end{feynhand}
			\end{tikzpicture}
		\end{minipage}
		\begin{minipage}{0.33\hsize}
			\centering
			\begin{tikzpicture}
				\begin{feynhand}
					\vertex[particle] (a) at (0,1.5) {$\Delta$};
					\vertex[particle] (b) at (-1.3,-0.75) {$\uc$};
					\vertex[particle] (c) at (1.3,-0.75) {$\uc$};
					\vertex[ringdot] (d) at (0,0) {};
					\propag[plain] (a) to (d);
					\propag[plain] (b) to (d);
					\propag[sca] (c) to (d);
				\end{feynhand}
			\end{tikzpicture}
		\end{minipage} \\
		\begin{minipage}{0.33\hsize}
			\centering
			\begin{tikzpicture}
				\begin{feynhand}
					\vertex[particle] (a) at (0,1.5) {$\Delta$};
					\vertex[particle] (b) at (-1.3,-0.75) {$\uc$};
					\vertex[particle] (c) at (1.3,-0.75) {$\uc$};
					\vertex[ringdot] (d) at (0,0) {};
					\propag[sca] (a) to (d);
					\propag[plain] (b) to (d);
					\propag[plain] (c) to (d);
				\end{feynhand}
			\end{tikzpicture}
		\end{minipage}
		\begin{minipage}{0.33\hsize}
			\centering
			\begin{tikzpicture}
				\begin{feynhand}
					\vertex[particle] (a) at (0,1.5) {$\Delta$};
					\vertex[particle] (b) at (-1.3,-0.75) {$\uc$};
					\vertex[particle] (c) at (1.3,-0.75) {$\uc$};
					\vertex[dot] (d) at (0,0) {};
					\propag[plain] (a) to (d);
					\propag[sca] (b) to (d);
					\propag[sca] (c) to (d);
				\end{feynhand}
			\end{tikzpicture}
		\end{minipage}
		\begin{minipage}{0.33\hsize}
			\centering
			\begin{tikzpicture}
				\begin{feynhand}
					\vertex[particle] (a) at (0,1.5) {$\Delta$};
					\vertex[particle] (b) at (-1.3,-0.75) {$\uc$};
					\vertex[particle] (c) at (1.3,-0.75) {$\uc$};
					\vertex[dot] (d) at (0,0) {};
					\propag[sca] (a) to (d);
					\propag[sca] (b) to (d);
					\propag[plain] (c) to (d);
				\end{feynhand}
			\end{tikzpicture}
		\end{minipage}
	\end{tabular}
	\caption{Main vertices associated with the cubic action~\eqref{eq: S3}. The vertex values are $ia^2\epsilon\eta'=ia^2\epsilon\pqty{\Delta\eta(\tau_\us)\delta(\tau-\tau_\us)+\Delta\eta(\tau_\ue)\delta(\tau-\tau_\ue)}$ (no dot), $-i\delta(\tau-\tilde{\tau})a^2\epsilon\eta$ (white dot), and $-2i\delta(\tau-\tilde{\tau})a\epsilon/H$ (black dot), respectively. $\Delta\eta(\tau_\us)=-\Delta\eta(\tau_\ue)=-6$ is the jump in the $\eta$ parameter, and $\tilde{\tau}$ is the time of the partner mode that is to be contracted with the $\Delta$ mode.}
	\label{fig: vertex}
\end{figure}
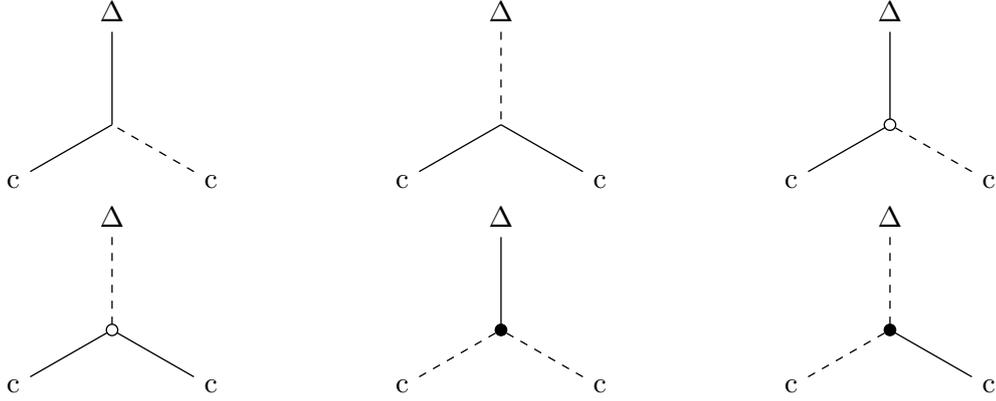

\begin{figure}
	\centering
	\begin{tabular}{c}
		\begin{minipage}{0.5\hsize}
			\centering
			\includegraphics[width=0.95\hsize]{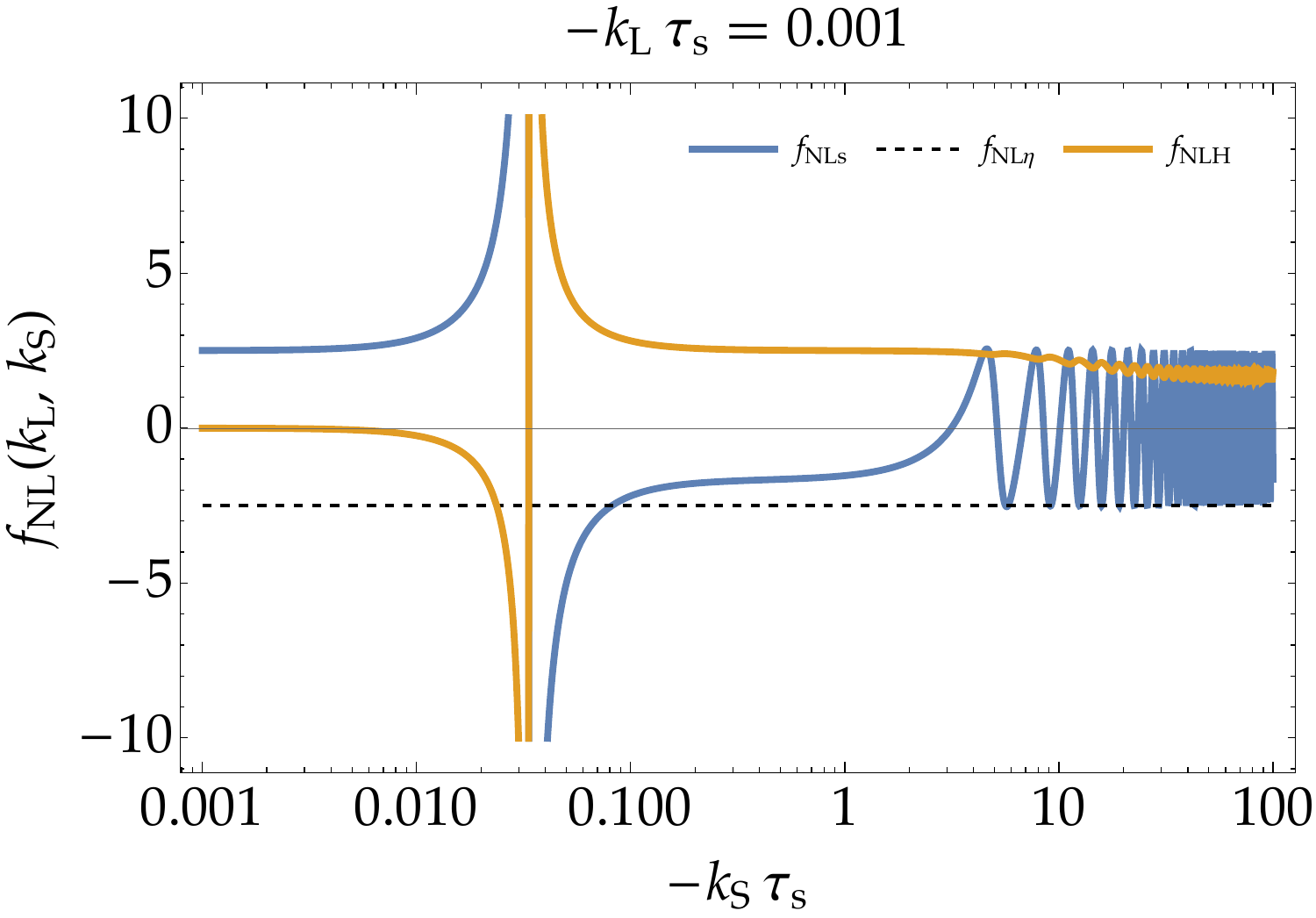}
		\end{minipage}
		\begin{minipage}{0.5\hsize}
			\centering
			\includegraphics[width=0.95\hsize]{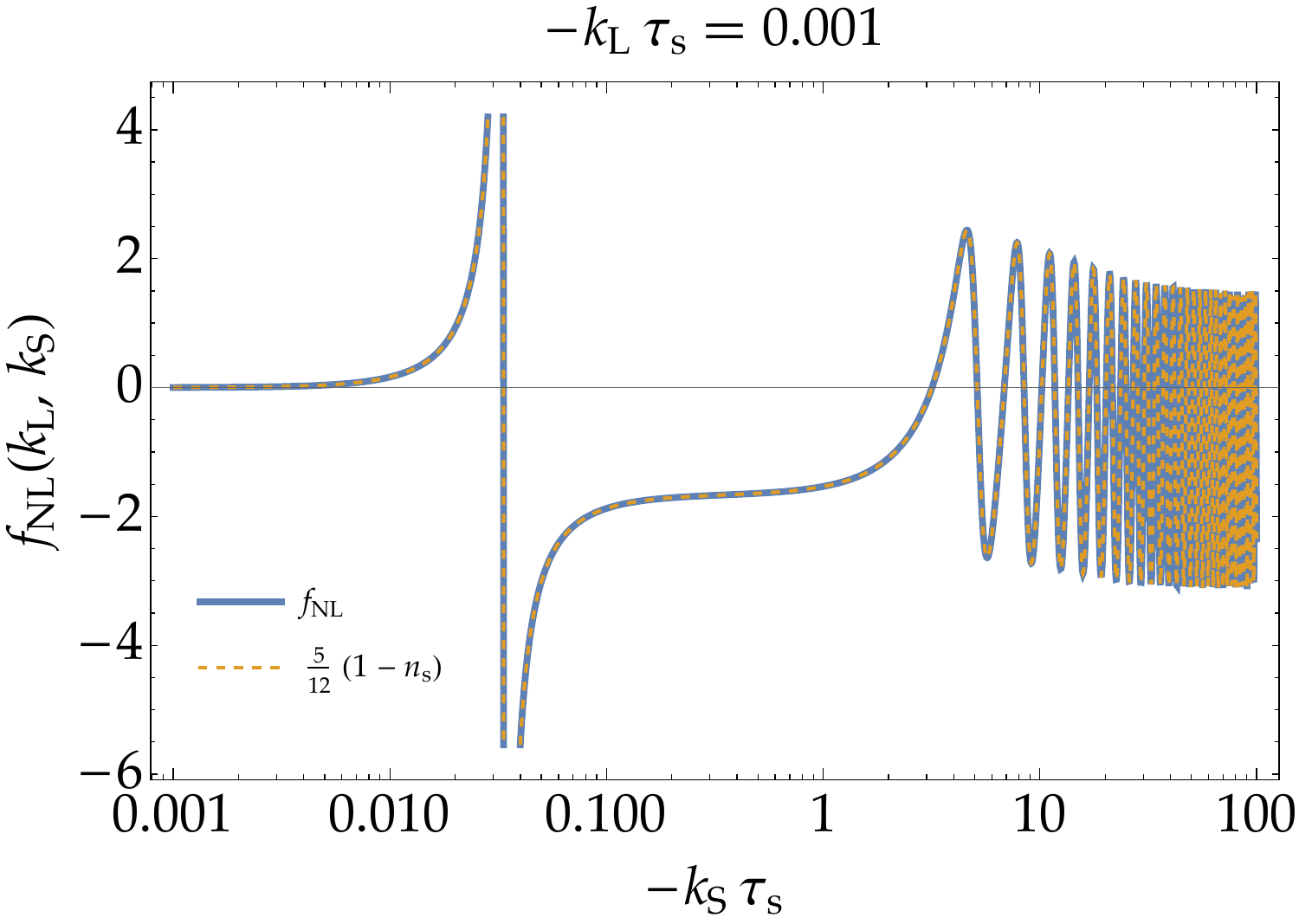}
		\end{minipage}
	\end{tabular}
	\caption{\emph{Left}: contributions from the first two diagrams (blue), the third one (black-dashed), and the fourth one (orange) in Eq.~\eqref{eq: bispectrum diagram USR} to $\fNL(k_\uL,k_\uS;\tau)$ at $\tau\to\tau_\ue-0$. \emph{Right}: a comparison between $\fNL$ (blue) and $\frac{5}{12}\pqty{1-\ns(k_\uS)}$ (orange-dashed) at the same time $\tau\to\tau_\ue-0$.}
	\label{fig: fNL USR}
\end{figure}

The cubic action relevant to the bispectrum in the transient \ac{USR} model is summarised as~\cite{Collins:2011mz,Arroja:2011yj,Adshead:2013zfa,Passaglia:2018afq,Passaglia:2018ixg}
\bae{\label{eq: S3}
	S^{(3)}\supset\int\dd{\tau}\dd[3]{\bmx}\bqty{\frac{a^2\epsilon}{2}\eta'\zeta^2\zeta'-\dv{\tau}\pqty{\frac{a^2\epsilon}{2}\eta\zeta^2\zeta'+\frac{a\epsilon}{H}\zeta{\zeta'}^2}},
}
where we have neglected higher-order terms in $\epsilon$ and terms that vanish by the equation of motion in the interaction picture as in Ref.~\cite{Fumagalli:2023hpa}.\footnote{The last term in Eq.~\eqref{eq: S3} was overlooked in Ref.~\cite{Fumagalli:2023hpa} though our Fig.~\ref{fig: fNL USR} shows its importance.}
This action leads to the vertices listed in Fig.~\ref{fig: vertex}.\footnote{The combinatorial factor arising at the Wick contraction is already included in the Feynman rules in the caption of Fig.~\ref{fig: vertex}.}
In the squeezed limit, the diagrams including the statistical propagator of the long mode $k_\uL$ dominate because $P_\zeta(k_\uL)$ is divergent as $\propto k_\uL^{-3}$ in the limit of $k_\uL\to0$.
Therefore, for the tree-level bispectrum calculation and the one-loop power spectrum calculation, the cubic vertices involving three $\zeta_\Delta$ and/or $\zeta'_\Delta$ (not shown in Fig.~\ref{fig: vertex}) give subleading-order (in $k_\text{L}^{-1}$) contributions, so we neglect these.

\begin{figure}
	\centering
	\begin{tabular}{c}
		\begin{minipage}{0.5\hsize}
			\centering
			\includegraphics[width=0.95\hsize]{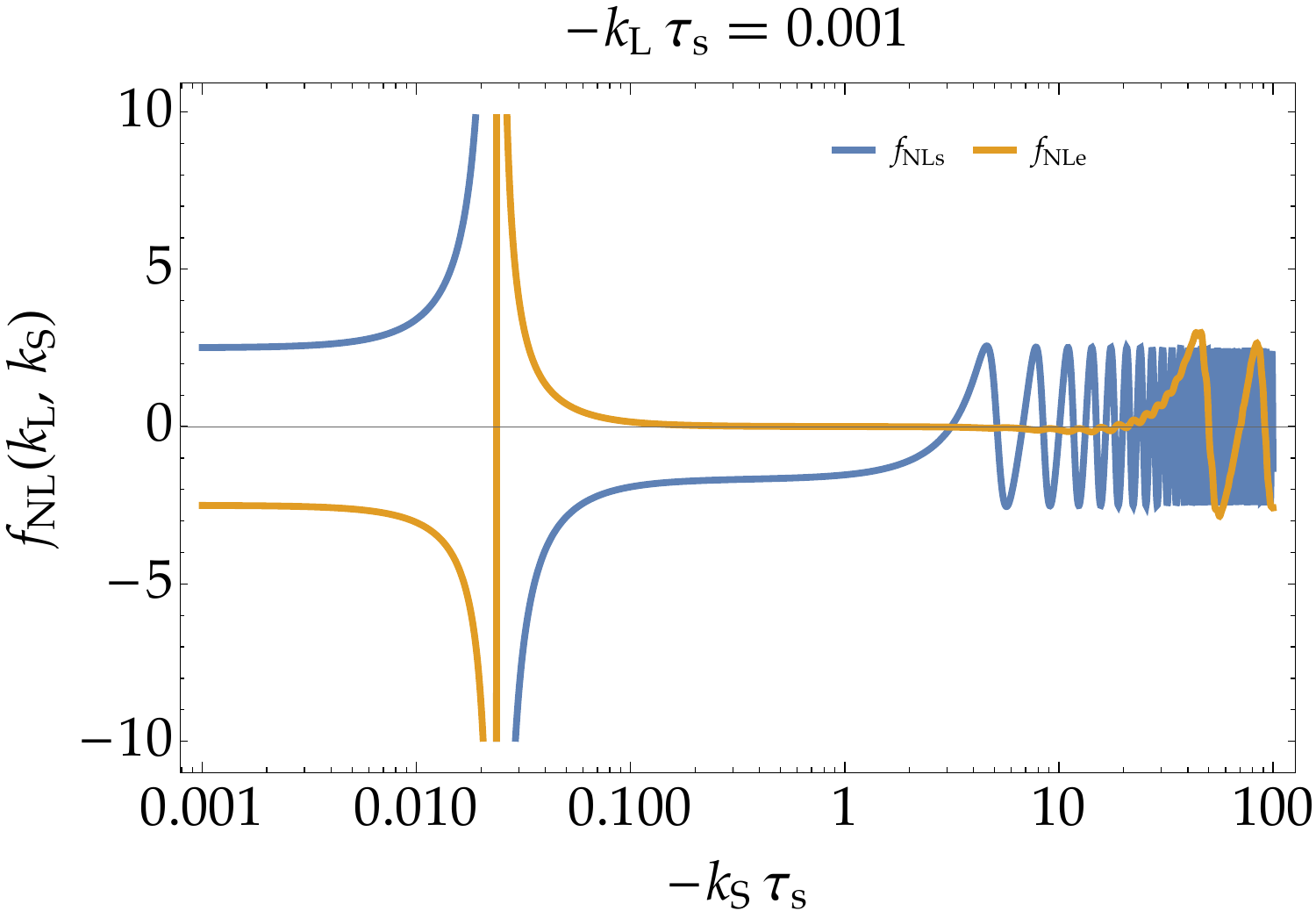}
		\end{minipage}
		\begin{minipage}{0.5\hsize}
			\centering
			\includegraphics[width=0.95\hsize]{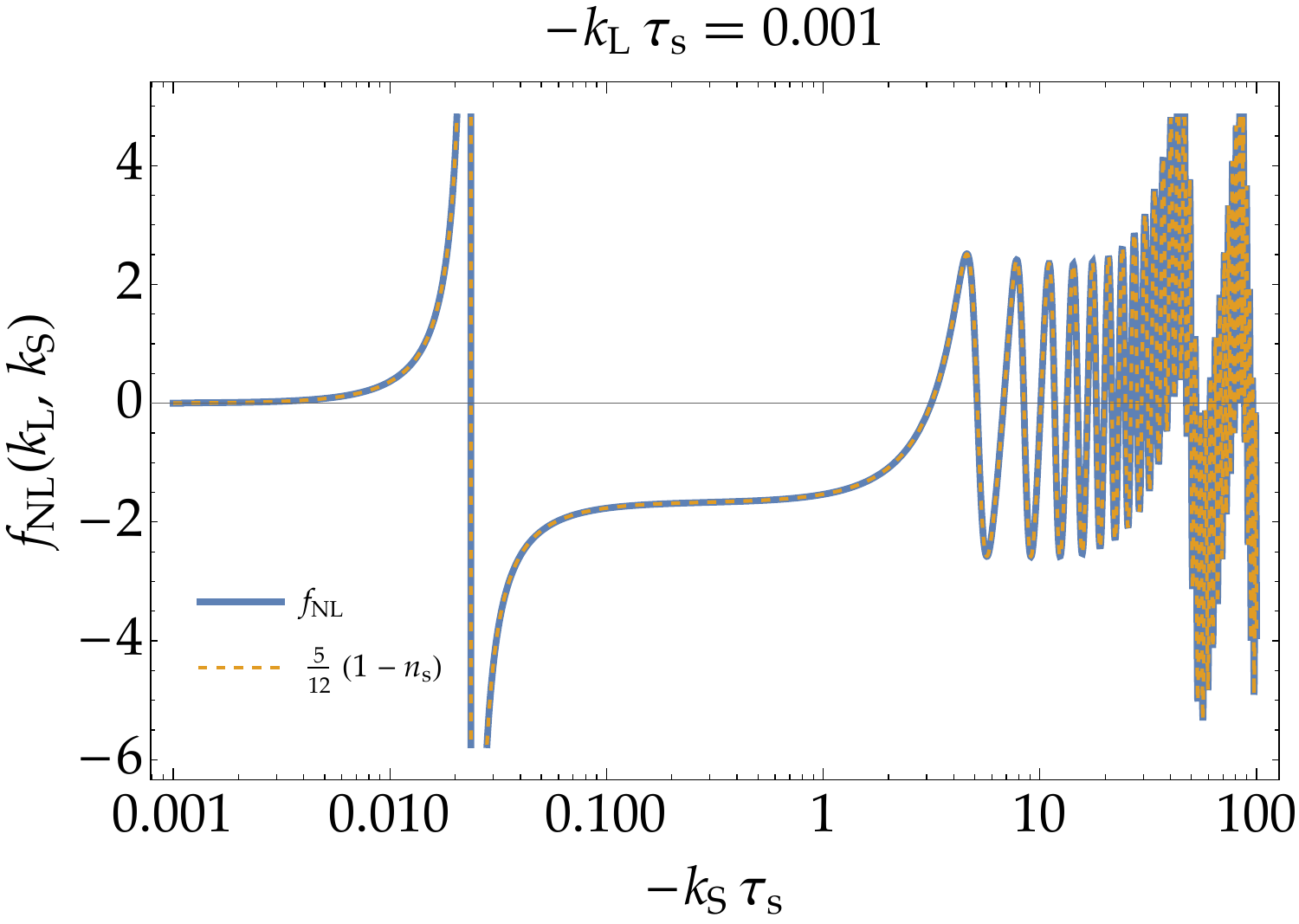}
		\end{minipage}
	\end{tabular}
	\caption{\emph{Left}: each contribution of $\tau_\us$ (blue) and $\tau_\ue$ (orange) to the generalised non-linearity parameter $\fNL(k_\uL,k_\uS;\tau)$~\eqref{eq: def of fNL} as functions of $k_\uS$ for $k_\uL=-0.001/\tau_\us$ and $\tau\to0$ with the same model parameters as Fig.~\ref{fig: calP and ns SUS}. \emph{Right}: a comparison between the total $\fNL$ (blue) and Maldacena's consistency relation $\frac{5}{12}\qty(1-\ns(k_\uS))$ (orange-dashed) at $\tau\to0$.} 
	\label{fig: fNL SR2}
\end{figure}

With use of these diagrams, one can calculate the bispectrum. As noted above, we focus on the leading-order contribution in the squeezed limit.
Noting that both $\eta$ and $\zeta'$ are suppressed in the \ac{SR} phase, one finds that the main diagrams for the squeezed bispectrum during the second \ac{SR} phase are given by
\bae{
	B_{\zeta\zeta\zeta}(k_1,k_2,k_3;\tau)&\eqqcolon
	\begin{tikzpicture}[baseline=(o.base)]
		\begin{feynhand}
			\vertex (a) at (-1.5,1) {\scriptsize $\zeta_\uc(\tau)$};
			\vertex (b) at (0.5,1) {\scriptsize $\zeta_\uc(\tau)$};
			\vertex (c) at (1.5,1) {\scriptsize $\zeta_\uc(\tau)$};
			\vertex[NEblob] (d) at (0,-1) {};
			\propag[plain] (d) to [out=140, in=-90, edge label=$k_1$] (a);
            \propag[plain] (d) to [edge label=$k_2$] (b);
			\propag[plain] (d) to [out=40, in=-90, edge label'=$k_3$] (c);
			\vertex (o) at (0,0);
		\end{feynhand}
	\end{tikzpicture} \nonumber \\
	\underset{\smqty{k_\uL\coloneqq k_1\ll k_2\simeq k_3\eqqcolon k_\uS, \\ \tau\gg\tau_\ue}}&{\simeq}
	\begin{tikzpicture}[baseline=(o.base)]
		\begin{feynhand}
			\vertex (a) at (-1.5,1);
			\vertex (b) at (0.5,1);
			\vertex (c) at (1.5,1);
			\vertex (d) at (0,-1);
            \vertex (e) at (0.25,0);
			\propag[plain] (d) to [out=135, in=-90, edge label=$k_1$] (a);
            \propag[sca] (d) to (e);
            \propag[plain] (e) to [edge label=$k_2$] (b);
			\propag[fermion] (d) to [out=45, in=-90, edge label'=$k_3$] (c);
			\vertex (o) at (0,0);
		\end{feynhand}
	\end{tikzpicture}
	+
	\begin{tikzpicture}[baseline=(o.base)]
		\begin{feynhand}
			\vertex (a) at (-1.5,1);
			\vertex (b) at (0.5,1);
			\vertex (c) at (1.5,1);
			\vertex (d) at (0,-1);
            \vertex (e) at (0.25,0);
			\propag[plain] (d) to [out=135, in=-90, edge label=$k_1$] (a);
            \propag[sca] (d) to (e);
			\propag[fermion] (e) to [edge label=$k_2$] (b);
			\propag[plain] (d) to [out=45, in=-90, edge label'=$k_3$] (c);
			\vertex (o) at (0,0);
		\end{feynhand}
	\end{tikzpicture} 
    + \, \text{($k_2\leftrightarrow k_3$)}.
}
Here and hereafter we neglect terms of order $(k_\uL/k_\uS)^2$, $(k_\uL\tau_\us)^2$, and $(k_\uL\tau_\ue)^2$.
If one fixes $k_\uL$ to the \ac{CMB} scale such that it is well frozen throughout the \ac{USR} and second \ac{SR} phase, one finds $G_{\uc\uc}(\tau,\tau_\us;k_\uL)\simeq G_{\uc\uc}(\tau,\tau_\ue;k_\uL)\simeq P_\zeta(\tau,k_\uL)$. The squeezed bispectrum is then summarised into the following form, 
\bae{\label{eq: Bz SR2}
	&B_{\zeta\zeta\zeta}(k_\uL,k_\uS,k_\uS;\tau) \nonumber \\
	&\bmbe{=\biggl(2ia^2(\tau_\us)\epsilon(\tau_\us)\Delta\eta(\tau_\us)\frac{G_{\uc\bar{\uc}}(\tau,\tau_\us;k_\uS)G_{\uc\Delta}(\tau,\tau_\us;k_\uS)+G_{\uc\bar{\Delta}}(\tau,\tau_\us;k_\uS)G_{\uc\uc}(\tau,\tau_\us;k_\uS)}{P_\zeta(\tau,k_\uS)} \\
	+\,\text{($\tau_\us\leftrightarrow\tau_\ue$)}\biggr)P_\zeta(\tau,k_\uL)P_\zeta(\tau,k_\uS).}
}
Defining the generalised non-linearity parameter $\fNL(k_\uL,k_\uS;\tau)$ by
\bae{\label{eq: def of fNL}
	\fNL(k_\uL,k_\uS;\tau)=\frac{5}{12}\frac{B_{\zeta\zeta\zeta}(k_\uL,k_\uS,k_\uS;\tau)}{P_\zeta(\tau,k_\uL)P_\zeta(\tau,k_\uS)},
}
the numerically calculated contribution of each term is presented in the left panel of Fig.~\ref{fig: fNL SR2}.
In the right panel, the total $\fNL$ is compared with $\frac{5}{12}\qty(1-\ns(k_\uS))$.
One finds that Maldacena's consistency relation completely holds.\footnote{The deviation from Maldacena's consistency relation cannot be seen even in the non-squeezed configuration, $k_\uL\sim k_\uS$, in Fig.~\ref{fig: fNL SR2}. This is because only terms dominant in the squeezed limit are kept in Eq.~\eqref{eq: Bz SR2}. The $\fNL$ parameter plotted in the right panel of Fig.~\ref{fig: fNL SR2} should be understood to be valid only in the squeezed limit, though the equilateral $\fNL$ is suppressed by slow-roll parameters anyway.}

The squeezed bispectrum during the \ac{USR} phase is also calculated similarly. The main diagrams are given by
\bme{\label{eq: bispectrum diagram USR}
	\begin{tikzpicture}[baseline=(o.base)]
		\begin{feynhand}
			\vertex (a) at (-1.5,1) {\scriptsize $\zeta_\uc(\tau)$};
			\vertex (b) at (0.5,1) {\scriptsize $\zeta_\uc(\tau)$};
			\vertex (c) at (1.5,1) {\scriptsize $\zeta_\uc(\tau)$};
			\vertex[NEblob] (d) at (0,-1) {};
			\propag[plain] (d) to [out=140, in=-90, edge label=$k_1$] (a);
            \propag[plain] (d) to [edge label=$k_2$] (b);
			\propag[plain] (d) to [out=40, in=-90, edge label'=$k_3$] (c);
			\vertex (o) at (0,0);
		\end{feynhand}
	\end{tikzpicture} 
	\underset{\smqty{k_\uL\coloneqq k_1\ll k_2\simeq k_3\eqqcolon k_\uS, \\ \tau_\us<\tau<\tau_\ue}}{\simeq}
	\begin{tikzpicture}[baseline=(o.base)]
		\begin{feynhand}
			\vertex (a) at (-1.5,1);
			\vertex (b) at (0.5,1);
			\vertex (c) at (1.5,1);
			\vertex (d) at (0,-1);
            \vertex (e) at (0.25,0);
			\propag[plain] (d) to [out=135, in=-90, edge label=$k_1$] (a);
            \propag[sca] (d) to (e);
            \propag[plain] (e) to [edge label=$k_2$] (b);
			\propag[fermion] (d) to [out=45, in=-90, edge label'=$k_3$] (c);
			\vertex (o) at (0,0);
		\end{feynhand}
	\end{tikzpicture}
	+
	\begin{tikzpicture}[baseline=(o.base)]
		\begin{feynhand}
			\vertex (a) at (-1.5,1);
			\vertex (b) at (0.5,1);
			\vertex (c) at (1.5,1);
			\vertex (d) at (0,-1);
            \vertex (e) at (0.25,0);
			\propag[plain] (d) to [out=135, in=-90, edge label=$k_1$] (a);
            \propag[sca] (d) to (e);
			\propag[fermion] (e) to [edge label=$k_2$] (b);
			\propag[plain] (d) to [out=45, in=-90, edge label'=$k_3$] (c);
			\vertex (o) at (0,0);
		\end{feynhand}
	\end{tikzpicture} \\
	+
	\begin{tikzpicture}[baseline=(o.base)]
		\begin{feynhand}
			\vertex (a) at (-1.5,1);
			\vertex (b) at (0.5,1);
			\vertex (c) at (1.5,1);
			\vertex[ringdot] (d) at (0,-1) {};
            \vertex (e) at (0.25,0);
			\propag[plain] (d) to [out=135, in=-90, edge label=$k_1$] (a);
            \propag[sca] (d) to (e);
			\propag[fermion] (e) to [edge label=$k_2$] (b);
			\propag[plain] (d) to [out=45, in=-90, edge label'=$k_3$] (c);
			\vertex (o) at (0,0);
		\end{feynhand}
	\end{tikzpicture}
	+
    \begin{tikzpicture}[baseline=(o.base)]
    	\begin{feynhand}
        	\vertex (a) at (-1.5,1);
			\vertex (b) at (0.5,1);
			\vertex (c) at (1.5,1);
			\vertex[dot] (d) at (0,-1) {};
            \vertex (e) at (0.25,0);
            \vertex (f) at (1,-0.2);
			\propag[plain] (d) to [out=135, in=-90, edge label=$k_1$] (a);
            \propag[sca] (d) to (e);
            \propag[fermion] (e) to [edge label=$k_2$] (b);
            \propag[sca] (d) to (f);
			\propag[plain] (f) to [out=45, in=-90, edge label'=$k_3$] (c);
			\vertex (o) at (0,0);
    	\end{feynhand}
    \end{tikzpicture}
    + \, \text{($k_2\leftrightarrow k_3$)}.
}
It is expressed as 
\bae{
	&B_{\zeta\zeta\zeta}(k_\uL,k_\uS,k_\uS;\tau) \nonumber \\
	&\bmbe{=\biggl(2ia^2(\tau_\us)\epsilon(\tau_\us)\Delta\eta(\tau_\us)\frac{G_{\uc\bar{\uc}}(\tau,\tau_\us;k_\uS)G_{\uc\Delta}(\tau,\tau_\us;k_\uS)+G_{\uc\bar{\Delta}}(\tau,\tau_\us;k_\uS)G_{\uc\uc}(\tau,\tau_\us;k_\uS)}{P_\zeta(\tau,k_\uS)} \\
	+\eta(\tau)-4i\frac{a(\tau)\epsilon(\tau)}{H(\tau)}\frac{G_{\uc\bar{\Delta}}(\tau,\tau;k_\uS)G_{\uc\bar{\uc}}(\tau,\tau;k_\uS)}{P_\zeta(\tau,k_\uS)}\biggr)P_\zeta(\tau,k_\uL)P_\zeta(\tau,k_\uS).}
}
Each contribution is plotted in the left panel of Fig.~\ref{fig: fNL USR} and the total $\fNL$ is compared with $\frac{5}{12}(1-\ns)$ in the right panel.
Maldacena's consistency relation is again satisfied. These facts can be summarised as the following important formula up to $\scrO\pqty{(k_\uL/k_\uS)^2, (k_\uL\tau_\us)^2, (k_\uL\tau_\ue)^2}$, 
\bae{\label{eq: soft theorem 1}
	\boxed{
	\begin{tikzpicture}[baseline=(o.base)]
		\begin{feynhand}
			\vertex (a) at (-1.5,1) {$\uc$};
			\vertex (b) at (0.5,1) {$\uc$};
			\vertex (c) at (1.5,1) {$\uc$};
			\vertex[NEblob] (d) at (0,-1) {};
			\propag[plain] (d) to [out=140, in=-90, edge label=$k_1$] (a);
            \propag[plain] (d) to [edge label=$k_2$] (b);
			\propag[plain] (d) to [out=40, in=-90, edge label'=$k_3$] (c);
			\vertex (o) at (0,0);
		\end{feynhand}
	\end{tikzpicture} 
	\underset{k_\uL=k_1\ll k_2\simeq k_3=k_\uS}{\simeq}-P_\zeta(k_\uL)P_\zeta(k_\uS)\dv{\ln\calP_\zeta(k_\uS)}{\ln k_\uS}=-\frac{2\pi^2}{k_\uS^3}P_\zeta(k_\uL)\dv{\calP_\zeta(k_\uS)}{\ln k_\uS}.}
}

Once one finds the expression of the squeezed bispectrum, one can also calculate three-point functions including the generalised momentum as
\bae{
	\braket{\hat{\zeta}_{\bmk_1}\hat{\zeta}_{\bmk_2}'\hat{\zeta}_{\bmk_3}}+\braket{\hat{\zeta}_{\bmk_1}\hat{\zeta}_{\bmk_2}\hat{\zeta}_{\bmk_3}'}\underset{k_1\ll aH}&{\simeq}\partial_\tau\braket{\hat{\zeta}_{\bmk_1}\hat{\zeta}_{\bmk_2}\hat{\zeta}_{\bmk_3}} \nonumber \\
	\underset{\smqty{k_\uL=k_1\ll k_2\simeq k_3=k_\uS, \\ k_\uL\ll aH}}&{\simeq}-(2\pi)^3\delta^{(3)}(\bmk_1+\bmk_2+\bmk_3)\frac{2\pi^2}{k_\uS^3}P_\zeta(k_\uL)\dv{\qty(\partial_\tau\calP_\zeta(k_\uL))}{\ln k_\uL},
}
where we neglected the time derivative of the $k_\uL$ mode as it is well frozen.
The following formula hence holds up to $\scrO\pqty{(k_\uL/k_\uS)^2, (k_\uL\tau_\us)^2, (k_\uL\tau_\ue)^2}$.
\bae{\label{eq: soft theorem 2}
	\boxed{
	\begin{tikzpicture}[baseline=(o.base)]
		\begin{feynhand}
			\vertex (a) at (-1.5,1) {$\uc$};
			\vertex (b) at (0.5,1) {$\uc$};
			\vertex (c) at (1.5,1) {$\uc$};
			\vertex[NEblob] (d) at (0,-1) {};
			\propag[plain] (d) to [out=140, in=-90, edge label=$k_1$] (a);
            \propag[sca] (d) to [edge label=$k_2$] (b);
			\propag[plain] (d) to [out=40, in=-90, edge label'=$k_3$] (c);
			\vertex (o) at (0,0);
		\end{feynhand}
	\end{tikzpicture} + \, \text{($k_2\leftrightarrow k_3$)}
	\underset{\smqty{k_\uL=k_1\ll k_2\simeq k_3=k_\uS, \\ k_\uL\ll aH}}{\simeq}-\frac{2\pi^2}{k_\uS^3}P_\zeta(k_\uL)\dv{\qty(\partial_\tau\calP_\zeta(k_\uL))}{\ln k_\uL}.
	}
}

\subsection{One-loop corrections to the power spectrum}

Armed with all the weapons, one can systematically calculate (cut-in-the-side) one-loop corrections on the soft two-point function.
Evaluated at a time well after the \ac{USR} phase, the (would-be) contributions which are not suppressed in the soft limit are summarised as
\bae{
    P_\zeta^{\text{(CIS)}}(\tau,k_\uL)&= P_\zeta^{\text{(CIS,1)}}(\tau,k_\uL)+P_\zeta^{\text{(CIS,2)}}(\tau,k_\uL)+P_\zeta^{\text{(CIS,3)}}(\tau,k_\uL) \nonumber \\
    &=\int^{\tau}\dd{\tau'}\int\frac{\dd[3]{\bmq}}{(2\pi)^3}\Biggl[
    \begin{tikzpicture}[baseline=(a.base)]
        \begin{feynhand}
            \vertex (a) at (-2,0);
            \vertex (b) at (2,0);
            \vertex[NEblob] (c) at (-0.75,0) {};
            \vertex (d) at (0.75,0);
            \vertex (dd) at (1,0.25) {$\tau'$};
            \vertex (e) at (1.25,0);
            \propag[plain] (a) to [edge label=$\bmk_\uL$] (c);
            \propag[plain] (c) to [out=45, in=90, edge label=$\bmq$] (d);
            \propag[plain] (c) to [out=-45, in=-90, edge label'=$\bmk_\uL-\bmq$] (d);
            \propag[sca] (d) to (e);
            \propag[fermion] (e) to [edge label=$\bmk_\uL$] (b);
        \end{feynhand}
    \end{tikzpicture} \nonumber \\
    &\qquad +\,
    \begin{tikzpicture}[baseline=(a.base)]
        \begin{feynhand}
            \vertex (a) at (-2,0);
            \vertex (b) at (2,0);
            \vertex[NEblob] (c) at (-0.75,0) {};
            \vertex (d) at (0.75,0);
            \vertex (dd) at (1,0.25) {$\tau'$};
            \propag[plain] (a) to [edge label=$\bmk_\uL$] (c);
            \propag[plain] (c) to [out=45, in=90, edge label=$\bmq$] (d);
            \propag[sca] (c) to [out=-45, in=-90, edge label'=$\bmk_\uL-\bmq$] (d);
            \propag[fermion] (d) to [edge label=$\quad\bmk_\uL$] (b);
        \end{feynhand}
    \end{tikzpicture}
    \,+\,
    \begin{tikzpicture}[baseline=(a.base)]
        \begin{feynhand}
            \vertex (a) at (-2,0);
            \vertex (b) at (2,0);
            \vertex[NEblob] (c) at (-0.75,0) {};
            \vertex[dot] (d) at (0.75,0) {};
            \vertex (dd) at (1,0.25) {$\tau'$};
            \vertex (e) at (1.25,0);
            \propag[plain] (a) to [edge label=$\bmk_\uL$] (c);
            \propag[plain] (c) to [out=45, in=90, edge label=$\bmq$] (d);
            \propag[sca] (c) to [out=-45, in=-90, edge label'=$\bmk_\uL-\bmq$] (d);
            \propag[sca] (d) to (e);
            \propag[fermion] (e) to [edge label=$\bmk_\uL$] (b);
        \end{feynhand}
    \end{tikzpicture}
    \Biggr].
    \label{eq:P_zeta_CIS}
}
Making use of the soft theorem~\eqref{eq: soft theorem 1}, the contribution of the first term, dubbed $P_\zeta^{\text{(CIS,1)}}$,\footnote{Strictly speaking, the bispectrum included in the diagram is not equal-time but the long-mode is at $\tau$ while the short-mode is at $\tau'\leq\tau$. However, since the long-mode of interest is well frozen throughout all important times, $\tau_\us$ and $\tau_\ue$, one can replace $\zeta_{k_\uL}(\tau)$ by $\zeta_{k_\uL}(\tau')$ and also $P_\zeta(\tau',k_\uL)$ by $P_\zeta(\tau,k_\uL)$.} reads
\bme{
    P_\zeta^\text{(CIS,1)}(\tau,k_\uL)=-ia^2(\tau_\us)\epsilon(\tau_\us)\Delta\eta(\tau_\us)G_{\uc\bar{\Delta}}(\tau,\tau_\us;k_\uL)P_\zeta(\tau,k_\uL)\int^{\ln k_\umax}_{\ln k_\umin}\dd{\ln q}\dv{\calP_\zeta(\tau_\us,q)}{\ln q} \\ 
    +\,\text{($\tau_\us\leftrightarrow\tau_\ue$)}.
}
The coefficient does not cause any singular behaviour as $G_{\uc\bar{\Delta}}(\tau,\tau_{\us/\ue};k_\uL)\sim i/(2a^2(\tau_{\us/\ue})\epsilon(\tau_{\us/\ue}))$ (see Appendix~\ref{sec:calculations}).
The $k_\umax\to\infty$ contribution can be dropped by the $i\varepsilon$ prescription as discussed in the previous section. 
Explicitly, 
\begin{align}
    \lim_{k_\text{max}\to \infty}\mathcal{P}_\zeta (\tau_\us, k_\text{max}) = \lim_{k_\text{max}\to \infty} \frac{ H^2}{8\pi^2 \epsilon_\text{SR}} \left((1 - \varepsilon k_\text{max} \tau_\us)^2 + k_\text{max}^2\tau_\us^2 \right)\ee^{2 \varepsilon k_\text{max} \tau_\text{s}}
    = 0.
\end{align}
A similar comment applies to $\mathcal{P}_\zeta (\tau_\text{e}, k_\text{max})$: it involves an exponential suppression factor $\ee^{2 \varepsilon k_\text{max} \tau_\text{e}} \to 0$ though the prefactor becomes more complicated.
We take $k_\umin$ as $k_\umin\gtrsim k_\uL$ to utilise the squeezed bispectrum\footnote{
Note that the requirement of the squeezed spectrum $k_\text{min} \gg k_\text{L}$ is not strict; Figs.~\ref{fig: fNL SR2} and \ref{fig: fNL USR} show that the consistency relation holds to a good precision even when $k_\text{min} \to k_\text{L}$. 
} but $k_\umin\ll k_\enh$ so that $\calP_\zeta(k_\umin)\ll\calP_\zeta(k_\enh)$, neglecting the contribution of $q<k_\umin$ which is expected to be small anyway.
After all, the ratio of the correction to the tree propagator,
\bae{
    \frac{P_\zeta^\text{(CIS,1)}(\tau,k_\uL)}{P_\zeta(\tau,k_\uL)}=-ia^2(\tau_\us)\epsilon(\tau_\us)\Delta\eta(\tau_\us)G_{\uc\bar{\Delta}}(\tau,\tau_\us;k_\uL)\calP_\zeta(\tau_\us,k_\umin)+\,\text{($\tau_\us\to\tau_\ue$)},
}
is negligibly small. For example, one can numerically evaluate it as $\sim\calO(10^{-10})$ in our setup for $\tau\to0$ and $-k_\uL\tau_\us=-k_\umin\tau_\us=10^{-3}$.

The second and third diagrams in eq.~\eqref{eq:P_zeta_CIS}, $P_\zeta^\text{(CIS,2)}$ and $P_\zeta^\text{(CIS,3)}$, are also negligible.
The soft theorem~\eqref{eq: soft theorem 2} leads to
\bae{
    &P_\zeta^\text{(CIS,2)}(\tau,k_\uL)+P_\zeta^\text{(CIS,3)}(\tau,k_\uL) \nonumber \\
    &=\bmte{-ia^2(\tau_\us)\epsilon(\tau_\us)\Delta\eta(\tau_\us)G_{\uc\Delta}(\tau,\tau_\us;k_\uL)P_\zeta(\tau,k_\uL)\int^{\ln k_\umax}_{\ln k_\umin}\dd{\ln q}\eval{\dv{\pqty{\partial_{\tau'}\calP_\zeta(\tau',q)}}{\ln q}}_{\tau'=\tau_\us} \\
    +\,\text{($\tau_\us\leftrightarrow\tau_\ue$)}\, 
    -\frac{P_\zeta(\tau,k_\uL)}{a(\tau)H(\tau)}\int^{\ln k_\umax}_{\ln k_\umin}\dd{\ln q}\dv{\pqty{\partial_{\tau}\calP_\zeta(\tau,q)}}{\ln q}} \nonumber \\
    \underset{k_\umax\to\infty}&{\to}\bmte{\Biggl[ia^2(\tau_\us)\epsilon(\tau_\us)\Delta\eta(\tau_\us)G_{\uc\Delta}(\tau,\tau_\us;k_\uL)\eval{\partial_{\tau'}\calP_\zeta(\tau',k_\umin)}_{\tau'=\tau_\us} \\
    +\,\text{($\tau_\us\leftrightarrow\tau_\ue$)}\,+\frac{\partial_\tau\calP_\zeta(\tau,k_\umin)}{a(\tau)H(\tau)}\Biggr]P_\zeta(\tau,k_\uL).}
}
The coefficients $-ia^2(\tau_{\us/\ue})\epsilon(\tau_{\us/\ue})G_{\uc\Delta}(\tau,\tau_{\us/\ue};k_\uL)$ are again independent of $k_\uL$ at the leading order and harmless (see Appendix~\ref{sec:calculations}). The summation of the first and second terms in the square brackets is numerically $\sim\calO(10^{-11})$ for $\tau\to0$ and $-k_\uL\tau_\us=-k_\umin\tau_\us=10^{-3}$ and it becomes even smaller for smaller $k_\uL$ and $k_\umin$ because $\partial_\tau\calP_\zeta(\tau,k)\propto k^2$.
The third term in the square brackets vanishes in the limit $\tau\to0$.

In the above calculations, it is crucial to cover the whole integration domain $[k_\text{min}, \, k_\text{max}] \to [k_\text{L}, \, \infty)$. In contrast, many authors in the recent discussions~\cite{Kristiano:2022maq, Riotto:2023hoz, Choudhury:2023jlt, Kristiano:2023scm, Riotto:2023gpm, Choudhury:2023rks, Firouzjahi:2023aum, Choudhury:2023hvf, Firouzjahi:2023ahg, Franciolini:2023lgy, Fumagalli:2023hpa, Maity:2023qzw} restricted the integration domain to $[k_\text{min}, \, k_\text{max}] \simeq [k_\text{s}, \, k_\text{e}]$, $[k_\text{L}, \, k_\text{e}]$, or similar ones by hand, where $k_\text{s}$ and $k_\text{e}$ are the wavenumbers that cross the Hubble horizon at $\tau = \tau_\text{s}$ and $\tau_\text{e}$, respectively. Although it may apparently make sense to extract the contribution only from the transient USR period, the above calculation shows that this introduces an artificially large contribution to the one-loop correction to the power spectrum of the soft modes of the curvature perturbations. As long as Maldacena's consistency relation holds (and we have confirmed it does hold), the integrand is written as the total derivative, so the large contribution is exactly cancelled by the adjacent integration domains. 

In summary, Maldacena's consistency relation as the cosmological soft theorem ensures the cancellation of the one-loop correction from small-scale perturbations. This is our main result.
It should be noted that we do not claim the one-loop correction vanishes on an arbitrary scale in the transient \ac{USR} model. For modes exiting the horizon right before the \ac{USR} onset $\tau_\us$, even though they are superhorizon throughout the \ac{USR} phase, they can grow at the tree level, Maldacena's consistency relation will be violated~\cite{Namjoo:2012aa}, and the one-loop correction can be sizable for them.
Self-loop corrections of the enhanced modes can also be non-negligible effects (see, e.g., Refs.~\cite{Abolhasani:2020xcg,Inomata:2022yte,Fumagalli:2023loc}).

\section{\boldmath Comment on the tadpole contributions and the stochastic-\texorpdfstring{$\delta N$}{Delta N} formalism}\label{sec: tadpole}

\begin{figure}
	\centering
	\begin{tabular}{c}
		\begin{minipage}{0.33\hsize}
			\centering
			\begin{tikzpicture}
				\begin{feynhand}
					\vertex[particle] (a) at (0,0.5) {$\uc$};
					\vertex[ringblob] (b) at (0,-0.5) {$\braket{\zeta}$};
					\propag[plain] (b) to (a);
				\end{feynhand}
			\end{tikzpicture}
		\end{minipage}
		\begin{minipage}{0.33\hsize}
			\centering
			\begin{tikzpicture}
				\begin{feynhand}	
					\vertex (a) at (-1.5,0);
					\vertex (b) at (1.5,0);
					\vertex[NEblob] (c) at (0,0) {};
					\vertex[ringblob] (d) at (0,1.5) {$\braket{\zeta}$};
					\propag[plain] (a) to [edge label'=$\bmk$] (c);
					\propag[plain] (b) to [edge label=$\bmk$] (c);
					\propag[plain] (d) to [edge label=\mbox{$\bmp=\bm{0}$}] (c);
				\end{feynhand}
			\end{tikzpicture}
		\end{minipage}
	\end{tabular}
    \caption{\emph{Left}: tadpole diagram $\braket{\zeta_\uc}$. \emph{Right}: tadpole correction on the two-point function.}
    \label{fig: tadpole}
\end{figure}
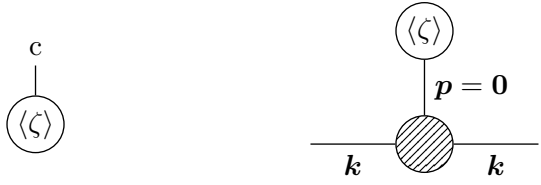

We comment on the tadpole contribution in this section.
Cubic interactions correct the two-point function not only by the ordinary self-energy diagrams shown in Fig.~\ref{fig: self energy} but also by the tadpole one illustrated in the right panel of Fig.~\ref{fig: tadpole} as the tadpole (left panel) is not necessarily prohibited for the curvature perturbation.
Non-vanishing tadpole is renormalised by the redefinition of the curvature perturbation, $\tilde{\zeta}=\zeta-\braket{\zeta}$. This redefinition does not change the correlation functions in the comoving Fourier space because $\tilde{\zeta}_{\bmk}=\zeta_{\bmk}$ for $k\neq0$.
However, it does change the relation between the comoving scale and the physical scale because it is equivalent to the redefinition of the global scale factor as $\tilde{a}=a\ee^{\braket{\zeta}}$, and hence one can fix the physical scale $k_\ph$ and define the renormalised (dimensionless) power spectrum $\calP_\zeta^\ren(k)$ so that the comoving scale $k$ corresponds to the physical scale of interest by $k_\ph=k/\tilde{a}$, that is,
\bae{
	\calP^\ren_\zeta(k):=\calP_\zeta(k\ee^{-\braket{\zeta}})=\calP_\zeta(k)-\braket{\zeta}\dv{\calP_\zeta(k)}{\ln k}+\calO(\braket{\zeta}^2),
}
or
\bae{
	P^\ren_\zeta(k)=P_\zeta(k)-\braket{\zeta}P_\zeta(k)\dv{\ln\calP_\zeta(k)}{\ln k}+\calO(\braket{\zeta}^2)\eqqcolon P_\zeta(k)+\Delta P_\zeta^\tad(k)+\calO(\braket{\zeta}^2).
}
The leading correction $\Delta P_\zeta^\tad(k)$ is suggestive to be understood as the right diagram of Fig.~\ref{fig: tadpole} by replacing the long-mode statistical propagator $G_{\uc\uc}(k_\uL)\simeq P_\zeta(k_\uL)$ by the tadpole graph $\braket{\zeta}$ (left panel of Fig.~\ref{fig: tadpole}) in the soft theorem~\eqref{eq: soft theorem 1}.\footnote{Strictly speaking, the replacement of $P_\zeta(k_\uL)$ by $\braket{\zeta}$ is not trivial because the soft theorem requires the constantcy of $P_\zeta(k_\uL)$ while $\braket{\zeta}$ may evolve during the \ac{USR} phase. 
The detailed proof is left for future work.}
It is proportional to the tree-level power $P_\zeta(k)$ and \ac{SR}-suppressed $\propto\dv*{\ln\calP_\zeta(k)}{\ln k}$ but would be non-negligible for very precise future observations.

Such a tadpole effect is also observed in the so-called \emph{stochastic-$\delta N$} approach~\cite{Fujita:2013cna,Fujita:2014tja,Vennin:2015hra,Ando:2020fjm,Tada:2021zzj}.
The stochastic formalism is known as an effective theory of the superhorizon-coarse-grained matter fields such as the inflaton.
There, the effective action for the coarse-grained fields is obtained by integrating out the subhorizon perturbations. It causes random Gaussian noise at the leading order, which can be understood as the horizon exit of perturbations.
Assuming that the coarse-grained fields are well-classicalised and follow the stationary point of the effective action, they behave as independent Brownian motions.
The inflatons' fluctuations are converted into the curvature perturbations by the $\delta N$ formalism~\cite{Salopek:1990jq,Sasaki:1995aw,Sasaki:1998ug,Lyth:2004gb}. The elapsed e-folding number $\calN$ can fluctuate due to the stochastic noise, even from the same initial condition to the same end-of-inflation condition, and such a fluctuation is nothing but the curvature perturbation according to the $\delta N$ formalism.
The (physical) scale $k_\ph$ of the power spectrum is then related to the field-space (or phase-space) points $N(k_\ph)=\ln(k_\uf/k_\ph)$ e-folds before the end of inflation, where $k_\uf\sim H_\uf$ is the coarse-graining scale at the end of inflation.
Here, even if the scale of interest corresponds to an attractor phase, if there is a very stochastic phase after that, the mean elapsed time $\braket{\calN}$ from a point can be shifted from $N_\cl$ estimated without noise.
This difference is understood as the tadpole $\braket{\zeta}=\braket{\calN}-N_\cl$ in the stochastic formalism and it causes the correction in the (dimensionless) power spectrum given by
\bae{
	(\braket{\calN}-N_\cl)\dv{\calP_\zeta}{N_\bw}=-\braket{\zeta}\dv{\calP_\zeta}{\ln k}=\frac{k^3}{2\pi^2}\Delta P_\zeta^\tad,
}
at the leading order in the attractor case. This is an interpretation of the tadpole correction in the stochastic approach.
We do not check the consistency in the tadpole value $\braket{\zeta}$ itself between the in-in and stochastic approach and leave it for future work as the stochastic calculation in the transient \ac{USR} model is a rich topic in itself.

Other than such a tadpole contribution, the stochastic approach would only predict the volume-suppressed ($(k_\uL/k_\uS)^3$-suppressed) correction, which is typical for the cut-in-the-middle ones.
Spatial coarse-graining of the curvature perturbations on a certain scale $k_\ph$ is implemented as a sampling average over paths branching at the time $N(k_\ph)$ e-folds before the end of inflation.
The large-scale perturbation generated by small-scale physics is hence understood as a deviation of the sampling average from the true average.
However, the central limit theorem guarantees that such a deviation is suppressed by the inverse sampling number, i.e., the volume ratio $(k_\uL/k_\uS)^3$.

\section{Comment on the tensor perturbations}\label{sec: tensor}

The loop correction on the tensor perturbations is also a hot topic. Ota et al.~\cite{Ota:2022hvh, Ota:2022xni} claimed that the superhorizon tensor power spectrum can be enhanced by the loop correction of a resonantly enhanced spectator scalar field. Firouzjahi~\cite{Firouzjahi:2023btw} computed the tensor loop correction from the inflaton in the transient \ac{USR} model and found that it is not significant but non-vanishing.

The soft tensor $h$ also satisfies Maldacena's consistency relation. There, the squeezed bispectrum between the soft tensor and two hard operators $\calO$ is given by the shear transformation of $\calO$'s power spectrum~\cite{Maldacena:2002vr,Pajer:2013ana}:
\bae{
	B_{h_\lambda\calO\calO}(k_1,k_2,k_3)\underset{k_1\ll k_2\sim k_3}{\to} - \frac{1}{2} P_{h_\lambda}(k_1)P_\calO(k_2)e^\lambda_{ij}(\hat{\bmk}_1)\hat{k}_2^i\hat{k}_2^j\dv{\ln P_\calO(k_2)}{\ln k_2},
}
where $\lambda$ labels the tensor polarisation and $e^\lambda_{ij}$ is the corresponding polarisation tensor, normalised by $e^{\lambda}_{ij}(\hat{\bmk})e^{\lambda'}_{ij}(\hat{\bmk})=\delta^{\lambda\lambda'}$.
However, this soft theorem does not ensure vanishing one-loop correction contrary to the scalar case.
Assuming that the other vertex brings another polarisation tensor $e^{\lambda'*}_{ij}(\hat{\bmk}_1)k_2^ik_2^j$ and making use of the solid angle integration $\int\dd{\hat{\bmk}_2}e^\lambda_{ij}(\hat{\bmk}_1)e^{\lambda'*}_{lm}(\hat{\bmk}_1)\hat{k}_2^i\hat{k}_2^j\hat{k}_2^l\hat{k}_2^m=\frac{8\pi}{15}\delta^{\lambda\lambda'}$, the one-loop correction from the coupling $h\calO\calO$ is expected to be proportional to
\bme{
	P_{h}(p)\int_{\ln k_\umin}^{\ln k_\umax}\frac{k^3}{2\pi^2}k^2P_\calO(k)\dv{\ln P_\calO(k)}{\ln k} \dd{\ln k} \\
	=P_{h}(p)\pqty{k_\umax^2\calP_\calO(k_\umax)-k_\umin^2\calP_\calO(k_\umin) } - 5P_h (p) \int_{\ln k_\umin}^{\ln k_\umax}k^2 \calP_\calO (k) \dd{\ln k}.
 \label{eq: tensor_oneloop_cubic}}
While the first term cancels the contribution of the enhanced power $\calP_\calO(k_\enh)$ similarly to the scalar case, the second one is affected by the enhanced power. It can be significant if the enhanced power is large enough. It would be interesting to evaluate the one-loop correction from the coupling $hh\calO\calO$ because it can be comparable to Eq.~\eqref{eq: tensor_oneloop_cubic}.

Note that the self-energy of gauge fields, i.e., vacuum polarisation, is renormalised into the gauge coupling in quantum field theory~\cite{Weinberg:1995mt}. 
The tensor loop correction would also be absorbed by the renormalisation of the gravitational coupling $G$. 
We leave this problem for future work, too.

\section{Conclusions}\label{sec: conclusion}

In this paper, we have studied the quantum corrections of the large-scale curvature perturbations from much smaller-scale modes beyond the standard single-clock inflation paradigm. We have taken into account surface terms in the action neglected in the literature and emphasised the role of Maldacena's consistency relation. 
We have numerically confirmed the consistency conditions with or without a time derivative of the curvature perturbations $\zeta'_k$ in the \ac{SR}/\ac{USR}/\ac{SR} scenario [Eqs.~\eqref{eq: soft theorem 1} and \eqref{eq: soft theorem 2}] both in the \ac{USR} regime and in the second \ac{SR} regime [Figs.~\ref{fig: fNL USR} and \ref{fig: fNL SR2}].  Once the consistency relation has been established in the \ac{SR}/\ac{USR}/\ac{SR} scenario, we have substituted the relation in the subdiagram of the one-loop correction to the power spectrum of the soft curvature perturbations.  As in the single-clock case~\cite{Pimentel:2012tw}, we have found that the one-loop corrections from the enhanced scale $k_\text{enh}$ cancel. We emphasise that this generally holds true beyond the \ac{SR}/\ac{USR}/\ac{SR} scenario we focused on as an example \emph{as long as Maldacena's consistency relation holds.}

Yet, our one-loop calculations are not complete since we have not included quartic interactions of $\zeta$. The quartic action of a related quantity $\delta \phi$ or $\zeta_n$~\cite{Maldacena:2002vr} can be found in Refs.~\cite{Jarnhus:2007ia, Dimastrogiovanni:2008af}. However, our calculation is based on the physical/original curvature perturbation $\zeta$, so we cannot use these results directly. Derivation of the quartic action of $\zeta$ including the surface terms is an important subject left for future work toward the complete proof of the absence of one-loop corrections from the small-scale physics. Despite the fact that it is not a proof, we believe our result strongly suggests that the small-scale loop effects do not significantly affect the curvature perturbations on much larger scales. 
 
Finally, let us discuss the implications of our results on the recent claim that primordial black hole production is excluded in single-field inflation scenarios. As we have seen, the quantum correction on a large scale such as the \ac{CMB} scale is insensitive to the \ac{USR} phase as long as these scales are hierarchically separated.  In our precision of calculation, we have been left with finite residual contributions but they are negligible compared to the tree-level contribution.  Thus, the one-loop correction is much smaller than the tree-level result, so there is no concern about the breakdown of perturbative loop expansion.  This is true even without the assumption of a smooth transition between either the \ac{SR} phase and the \ac{USR} phase. As we emphasised above, the same discussion applies to any \ac{PBH} production scenarios utilising the small-scale enhanced curvature perturbations that satisfy Maldacena's consistency relation. Therefore, \ac{PBH} production in single-field inflation scenarios is not excluded.

\acknowledgments

We are grateful to 
Ryo Saito
for helpful discussions.
Y.T. is supported by JSPS KAKENHI Grant
No.~JP21K13918.
This work was supported by IBS under the project code, IBS-R018-D1.

\appendix

\section{Retarded Green's function on superhorizon scales}\label{sec:calculations}

Here we show explicit expressions of $G_\text{c$\Delta$}(\tau,\tau';k_\uL)$ at the leading order when $-k_\uL\tau'\ll1$. 
When the condition i) $\tau,\tau'>\tau_\ue$, ii) $\tau_\us<\tau,\tau'<\tau_\ue$, or iii) $\tau,\tau'<\tau_\us$ is satisfied, $G_\text{c$\Delta$}(\tau,\tau';k_\uL)$ describes the time evolution of given mode $k_\uL$ during the single \ac{SR} or \ac{USR} phase. Accordingly, we have a simple expression at the leading order for such $(\tau,\tau')$ as
\bae{\label{eq:ret_single}
    G_\text{c$\Delta$}(\tau,\tau';k_\uL)
    &=
    -\frac{iH^2}{6\sqrt{\epsilon(\tau)\epsilon(\tau')}}
    \left(
        \tau^3-\tau'^3
    \right) \Theta(\tau - \tau')
    +\mathcal{O}\qty((k_\uL\tau')^2) \nonumber \\
    &= \frac{i}{6 H \sqrt{\epsilon(\tau)\epsilon(\tau')}}
    \left(
        \frac{1}{a^3(\tau)} - \frac{1}{a^3(\tau')}
    \right)  \Theta(\tau - \tau')
    +\mathcal{O}\qty((k_\uL\tau')^2).
}
When there is a transition from \ac{SR} to \ac{USR} or \ac{USR} to \ac{SR} between $\tau'$ and $\tau$, Eq.~\eqref{eq:ret_single} is no longer valid. For such $(\tau,\tau')$, we can use the following relation to evaluate $G_\text{c$\Delta$}(\tau,\tau';k_\uL)$ up to negligible $\mathcal{O}\qty((k_\uL\tau')^2)$ terms, 
\begin{align}
    G_\text{c$\Delta$}(\tau,\tau';k_\uL)
    =
    \begin{cases}
         G_\text{c$\Delta$}(\tau,\tau_\ue;k_\uL)
         +
         G_\text{c$\Delta$}(\tau_\ue,\tau_\us;k_\uL)
         +
         G_\text{c$\Delta$}(\tau_\us,\tau';k_\uL), & \tau>\tau_\ue>\tau_\us>\tau', \\
         G_\text{c$\Delta$}(\tau,\tau_\ue;k_\uL)
         +
         G_\text{c$\Delta$}(
         \tau_\ue,\tau';k_\uL), & \tau>\tau_\ue>\tau'>\tau_\us, \\
         G_\text{c$\Delta$}(\tau,\tau_\us;k_\uL)
         +
         G_\text{c$\Delta$}(\tau_\us,\tau';k_\uL), & \tau_\ue>\tau>\tau_\us>\tau'. 
    \end{cases}
    \label{eq:ret_transition}
\end{align}
We can show the validity of Eq.~\eqref{eq:ret_transition} by explicit calculations. Each term on the right-hand side of Eq.~\eqref{eq:ret_transition} can be evaluated by using Eq.~\eqref{eq:ret_single}. Eqs.~\eqref{eq:ret_single} and \eqref{eq:ret_transition} show that $G_\text{c$\Delta$}(\tau,\tau';k_\uL)$ is independent of $k_\uL$ at the leading order. Note that Eq.~\eqref{eq:ret_transition} also follows from the intuition that each coarse-grained patch on a superhorizon scale evolves in time independently at the leading order. 
In particular, for $\tau>\tau_\ue$, one finds
\beae{
    &G_\text{c$\Delta$}(\tau,\tau_\text{e};k_\uL)
    = \frac{-i}{6 H a^3(\tau_\text{e}) \epsilon (\tau_\text{e})} \left( 1- \left( \frac{a(\tau_\text{e})}{a(\tau)} \right)^3 \right)
    +\mathcal{O}\qty((k_\uL\tau_\text{e})^2), \\
    &G_\text{c$\Delta$}(\tau,\tau_\text{s};k_\uL)
    = \frac{-i}{6 H a^3(\tau_\text{s}) \epsilon (\tau_\text{s})} \left( \frac{a(\tau_\text{e})}{a(\tau_\text{s})} \right)^3 \left( 2 - \left(\frac{a(\tau_\text{e})}{a(\tau)}\right)^3 - \left(\frac{a(\tau_\text{s})}{a(\tau_\text{e})}\right)^3 \right) +\mathcal{O}\qty((k_\uL\tau_\text{s})^2).
}

Similarly, we obtain 
\begin{align}
    G_{\uc\bar{\Delta}}(\tau, \tau'; k_\text{L}) &= \frac{i H^2}{2 \sqrt{\epsilon(\tau)\epsilon(\tau')}} \left( \tau'{}^2 - \frac{\eta(\tau')}{6 \tau'}\left(\tau^3 - \tau'^3 \right) \right) \Theta(\tau - \tau')+\mathcal{O}\qty((k_\uL\tau')^2) \nonumber \\
    &= \frac{i}{2 a^2 (\tau') \sqrt{\epsilon(\tau) \epsilon(\tau')}} \left(1 + \frac{\eta(\tau')}{6} \left(1 - \left( \frac{a(\tau')}{a(\tau)} \right)^3 \right) \right) \Theta(\tau - \tau')+\mathcal{O}\qty((k_\uL\tau')^2),
\end{align}
in the case if both $\tau$ and $\tau'$ are either in the first \ac{SR}, the transient \ac{USR}, or in the second \ac{SR} phase. The analogy of Eq.~\eqref{eq:ret_transition} (just replacing $\Delta$ with $\bar{\Delta}$) does not hold. In the main text, we are interested in specific cases with $\tau' = \tau_\text{s}$ and $\tau_\text{e}$ with $\tau$ in the second \ac{SR} phase. In this case, 
\begin{align}
    G_{\uc\bar{\Delta}}(\tau, \tau_\text{s/e}; k_\text{L}) =& \frac{i}{2 a^2(\tau_\text{s/e} ) \epsilon(\tau_\text{s/e})} +\mathcal{O}\qty((k_\uL\tau_\text{s/e})^2).
\end{align}

\bibliographystyle{JHEP}
\bibliography{draft}
\end{document}